\documentclass[lettersize,journal]{IEEEtran}
\usepackage{amsmath,amsfonts}
\usepackage{algorithmic}
\usepackage{array}
\usepackage[caption=false,font=normalsize,labelfont=sf,textfont=sf]{subfig}
\usepackage{textcomp}
\usepackage{stfloats}
\usepackage{url}
\usepackage{verbatim}
\usepackage{graphicx}
\def\BibTeX{{\rm B\kern-.05em{\sc i\kern-.025em b}\kern-.08em
    T\kern-.1667em\lower.7ex\hbox{E}\kern-.125emX}}
\usepackage{balance}

\usepackage{gav_style}   

\begin{document}
\title{A Generative Diffusion Model for Probabilistic Ensembles of Precipitation Maps Conditioned on Multisensor Satellite Observations}
\author{Cl\'ement Guilloteau, Gavin Kerrigan, Kai Nelson, Giosue Migliorini, Padhraic Smyth, Runze Li, and Efi Foufoula-Georgiou
\thanks{Cl\'ement Guilloteau, Runze Li and Efi Foufoula-Georgiou are with the Department of Civil and Environmental Engineering at the University of California Irvine. Gavin Kerrigan, Kai Nelson and Padhraic Smyth are with the Department of Computer Science at the University of California Irvine. Giosue Migliorini is with the Department of Statistics at the University of California Irvine.}}

\markboth{This is a preprint, this research is currently under review for publication in IEEE TGRS}%
{How to Use the IEEEtran \LaTeX \ Templates}

\maketitle

\begin{abstract}
A generative diffusion model is used to produce probabilistic ensembles of precipitation intensity maps at the 1-hour 5-km resolution. The generation is conditioned on infrared and microwave radiometric measurements from the GOES and DMSP satellites and is trained with merged ground radar and gauge data over southeastern United States. The generated precipitation maps reproduce the spatial autocovariance and other multiscale statistical properties of the gauge-radar reference fields on average. Conditioning the generation on the satellite measurements allows us to constrain the magnitude and location of each generated precipitation feature. The mean of the 128-member ensemble shows high spatial coherence with the reference fields with 0.82 linear correlation between the two. On average, the coherence between any two ensemble members is approximately the same as the coherence between  any ensemble member and the ground reference, attesting that the ensemble dispersion is a proper measure of the estimation uncertainty. From the generated ensembles we can easily derive the probability of the precipitation intensity exceeding any given intensity threshold, at the 5-km resolution of the generation or at any desired aggregated resolution.
\end{abstract}


\section{Introduction}

\noindent The estimation of precipitation intensity from spaceborne passive radiometric measurements, either in the optical infrared (IR) domain or in the microwave (MW) domain has been performed operationally for more than two decades \cite{huffman1997GPCP,kidd2011status, kidd2021theglobal}. It is the backbone of a suite of global precipitation mapping products \cite{Xie2017CMORPH,Nguyen2018PERSIANN,huffman2020IMERG,kubota2020GSmAP}, which are widely used for research and applications in hydrology, water management, weather and climate monitoring, impact studies and risk assessment. In recent years, there has been increasing interest in applying deep learning methods to the problem of inverting the radiometric measurements into precipitation intensity \cite{Sano2015PNPR,Sadeghi2019PERSIANNcnn,Gorooh2022deepSTEP,Pfreundschuh2022GPROFNN,Guilloteau2023Constraining}. One of the main advantages of the deep learning framework is that it can easily accommodate a large number of heterogeneous inputs (predictor variables or conditioning variables) as constraints toward the output of a predictive or a conditional generative model. One can thus easily combine measurements from several different instruments as inputs to a deep learning model, allowing researchers to build relatively complex prediction models in a relatively straightforward data-driven manner. 

Importantly, deep neural networks are resilient to what is commonly referred to as the “curse of dimensionality”, which is the fact that the efficiency of classical estimation algorithms tends to dramatically decrease if too many variables are added as inputs \cite{Marimont1979curse,Poggio2017curse}. This “curse of dimensionality” can be viewed as a manifestation of the overfitting phenomenon: when empirically fitting an explicit statistical model with $n$ parameters ($n$ degrees of freedom) to a dataset, the number of independent samples necessary to avoid overfitting increases exponentially with $n$. However, within the deep learning framework, there is a-priori no contraindication to including a large number of predictors or conditions as input of a deep neural network for estimating precipitation intensity, with the objective of reinforcing the constraints and reducing the underdetermination. For these reasons, deep learning approaches are increasingly being applied to the problem of estimating precipitation intensity from coincident radiometric measurements made by sensors of different types and carried by different satellite platforms.

While these new-generation algorithms have demonstrated marked improvement in terms of retrieval accuracy and computational efficiency when compared to older methods such as lookup tables (LUT) or k-nearest neighbors (KNN) algorithms \cite{Pfreundschuh2024v7andbeyond}, the inversion of the measured radiances into precipitation intensity remains a fundamentally underconstrained problem \cite{Guilloteau2020beyond}. As a result, the precipitation intensities estimated from the radiometric measurements are always associated with some degree of uncertainty, and should be interpreted in a statistical or probabilistic sense \cite{Kirstetter2018probalistic}. Yet, many operational precipitation products such as \cite{Xie2017CMORPH,Nguyen2018PERSIANN,huffman2020IMERG,kubota2020GSmAP} do  not systematically provide a quantitative measure of the uncertainty associated with each estimate.

In the present study we introduce a deep neural conditional generative diffusion model for mapping precipitation intensity at the hourly time scale from coincident MW and IR satellite measurements. The MW radiances at the top of the atmosphere are provided by the SSMI/S imager onboard the DMSP satellite series, and the IR cloud brightness temperature at 10.3 µm is provided by the Advanced Baseline Imager (ABI) onboard the GOES-R satellite series. Most existing radiance-to-precipitation-intensity inversion algorithms focus on estimating the “instantaneous” precipitation intensity at the exact time of the radiometric measurement. Here we estimate the mean precipitation intensity for the one-hour period that spans from 30 min before the time of the SSMI/S MW radiometric measurement to 30 min after the time of the SSMI/S measurement. The south-eastern part of the continental United States is selected as a testbed area for the development and evaluation of the algorithm. Over this region, the ABI imager onboard GOES-16 provides one radiometric image every 5 minutes. For an SSMI/S observation occurring at time $t$, corresponding to the overpass time of the DMPS-F17 satellite over the region of interest, we take the 10-channel multispectral image produced by SSMI/S along with the 13 IR 10.3 µm images produced by ABI from $t-30min$ to $t+30min$ as inputs of the neural network. The neural network diffusion model generates a map of the hourly precipitation intensity from $t-30min$ to $t+30min$ within the area covered by the SSMI/S scan, for every overpass of the DMPS-F17 satellite. The target spatial resolution is 5 km. For training and evaluating the model, we rely on precipitation maps provided by the NOAA Multi-Radar Multi-Sensor (MRMS) system which combines ground radar and gauge measurements.

To account for the uncertainty in the generated fields of precipitation intensity, which is inherent to the limited information content of the passive radiometric measurements and the underdetermination of the inversion, we use the stochastic generation capabilities of the diffusion model to produce a large ensemble of possible realizations (128 realizations in the retained setup) of the precipitation field for each observed scene, instead of a unique “best guess”. All realizations are considered equiprobable, and the ensemble dispersion is treated as a measure of the retrieval uncertainty. While, over a large number of scenes, the ensemble mean shall be on average the closest estimate to the truth (in terms of the mean squared error), each realization is a “realistic” precipitation field in the sense that it preserves the expected statistical properties of the true precipitation field (in particular the statistical distribution of the pixel-wise intensities and the spatial correlation). The proposed technique is designated as DifERS (Diffusion-based Ensemble Rainfall estimation from Satellite).      

The two principal novelties of DifERS thus are: 1 - the handling of the uncertainty through the generation of ensembles of equiprobable realizations; 2 - the use of coincident measurements from different instruments and different platforms at the radiance inversion level. We note that, for the current operational multisensor/multiplatform satellite precipitation products, the fusion between the sensors is not performed at the inversion step, but is done in a subsequent step: the radiometric measurements of each sensor are independently inverted into precipitation intensity estimates, the different intensity estimates are then merged together through weighted averaging and dynamical interpolation \cite{huffman2020IMERG,Ushio2009GSMAPmvk,Joyce2011kalman}. Over both land and oceans, estimates of instantaneous precipitation intensity derived from passive MW measurements are generally found to be significantly more accurate than those derived from IR or multispectral optical imagery \cite{kidd2021theglobal}; consequently, more weight is given to MW estimates on average in multisensor products such as \cite{huffman2020IMERG}. We also note that, while with existing algorithms the radiometric measurements are generally inverted into precipitation intensities as “instantaneous snapshots”, it is often assumed that this snapshot is representative of the time-averaged intensity over periods ranging from 30 min to several hours at the later steps when the individual estimates are merged together \cite{kidd2021theglobal,Turk2009validating}. The short-time variations of precipitation intensity are thus ignored. Here, we specifically train DifERS to retrieve hourly-accumulated precipitation fields instead of instantaneous intensities as we expect that the 5-min IR imagery can help reduce the inaccuracy arising from variations of the instantaneous precipitation intensity during the hour of interest.

The article is organized as follows: Section 2 presents the data, Section 3 describes the architecture of the DifERS model and its training process, Section 4 presents the results in terms of the fidelity of the generated precipitation fields, measures of accuracy and uncertainty, and probabilistic interpretation of the DifERS ensembles. Finally, conclusions and future perspectives are presented in Section 5. 

\section{Data}
\begin{figure*}
    \centering
    \includegraphics[width=.8\linewidth]{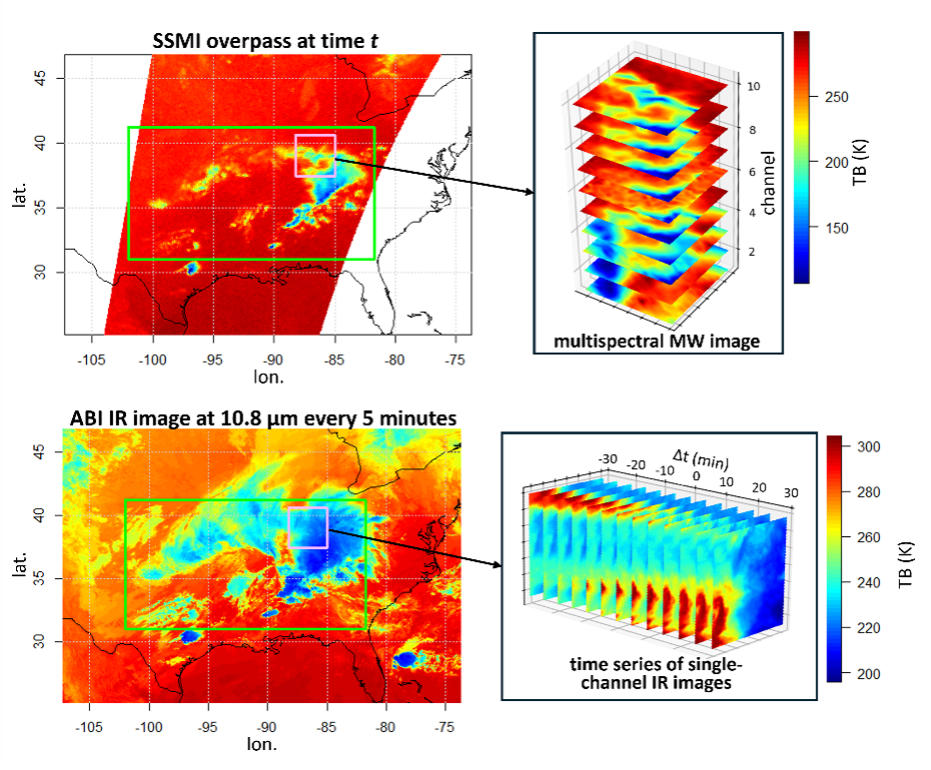}
    \caption{Passive MW and IR brightness temperatures measured at the top of the atmosphere on 2021-05-04 at 13:10 UTC over the study domain. (top, left) Brightness temperature at 92 GHz from SSMI/S onboard DMSP-F17. (top, right) Stacked SSMI/S brightness temperature for all channels over a 320 km by 320 km subset of the study domain. (bottom left) Brightness temperature at 10.3 µm from ABI onboard GOES-16 at 13:10 UTC (corresponding to the time of overpass of the DMSP-F17 satellite). (bottom right) Time series of ABI brightness temperature fields at 10.3 µm from 12:40 UTC to 13:40 UTC over a 320 km by 320 km subset of the study domain. The green rectangle on the left panel delineates the study domain between latitudes 31°N and 41°N and longitudes 81°W and 102°W.}
    \label{fig:1}
\end{figure*}
\subsection{Passive multispectral microwave radiances from SSMI/S}

\noindent The Special Sensor Microwave Imager/Sounder (SSMI/S) is a conically scanning passive microwave imager carried by the Defense Meteorological Satellite Program (DMSP) satellite series since 2003 \cite{Kunkee2008SSMIS}. The 24 channels of SSMI/S measure polarized microwave radiances at the top of the atmosphere between 19 and 183 GHz. Of the 24 channels, 11 are relevant for precipitation estimation: three vertical polarization and three horizontal polarization channels at 19, 37 and 92 GHz, a vertical polarization channel at 22 GHz, a horizontal polarization channel at 155 GHz, and three double-sideband vertical polarization channels at 183$\pm$1, 183$\pm$3 and 183$\pm$7 GHz. The radiances measured by SSMI/S at the top of the atmosphere are the product of the surface emission combined with the absorption and emission signal of water vapor and liquid rain drops in the atmosphere (predominant below 40 GHz), plus the scattering effect of atmospheric ice particles (predominant above 40 GHz). The conical scanning pattern of SSMI/S from an altitude of about 830 km covers a 1700-km wide swath under the satellite track. Each scan line consists of  180 overlapping fields of view and the distance between two consecutive scan lines is 12.5 km. While the local observation time is variable, a single SSMI/S sensor revisits any point of the globe at least twice per day. The current constellation of three functional SSMI/S sensors (onboard the F-16, F-17 and F-18 DMSP satellites) therefore allows for six observations per day at any location. For the algorithmic demonstration of the present study, only data provided by the SSMI/S sensor onboard DMSP F-17 is used. We note that since April 2016 the data from the 37V channel of the DMSP F-17 SSMI/S sensor is no longer available \cite{NASAF17}. Therefore, only the 10 remaining channels are used here. The DMSP F17 SSMI/S radiometric fields are distributed by NASA as the GPM\_1CF17SSMIS product (doi: 10.5067/GPM/SSMIS/F17/1C/07).

\subsection{Cloud brightness temperature at 10.3 $\mu$m}

\noindent The Advanced Baseline Imager (ABI) onboard the GOES-R satellite series performs optical imaging of the Earth from geostationary orbit in the visible and IR domain. In the present study we only consider the thermal IR 10.3 µm channel. The brightness temperature at 10.3 $\mu$m is a fair indicator of the altitude of the cloud top \cite{Heidinger2020ABI}. Because optical wavelengths cannot penetrate optically thick clouds, IR measurements provide only indirect information about the precipitation processes. The utility of geostationary optical imagery, compared to MW radiometry from low Earth orbit, lies in its’ ability to produce images frequently and at high spatial resolution. Over the continental United States the ABI imager onboard GOES-16 produces one image every five minutes at 2 km resolution. This high temporal sampling allows monitoring of the development and dynamics of cloud systems. The dynamical information provided by the ABI over a one-hour time window (from $t-30min$ to $t+30min$) complements the snapshot information provided by SSMI/S at time $t$. In \cite{Guilloteau2024lifecycle} it was demonstrated for example that the development stage of cloud systems can be estimated from geostationary IR, and provides useful information for reducing systematic biases in passive MW estimates of precipitation intensity. The ABI brightness temperature fields are distributed by NOAA as the ABI Level 2 Cloud and Moisture Imagery products (ABI-L2-CMIP).

\subsection{Surface precipitation intensity from MRMS ground radar and gauge measurements}

\noindent The reference precipitation fields used for training and evaluating DifERS are produced by the NOAA Multi-Radar Multi-Sensor (MRMS) system. The MRMS precipitation estimates are obtained from merged ground radar and gauge data \cite{Zhang2016MRMS}. They are produced at a 1 km and 2 minutes resolution over the contiguous United States and the southern part of Canada. The present study is focused over the southeastern part of the United States, between latitudes 31$^\circ$N and 41$^\circ$N and longitudes 81$^\circ$W and 102$^\circ$W (Figure \ref{fig:1}), where the radar coverage is spatially continuous and the gauge density is high. \\

All the radiometric images from SSMI/S and ABI are projected on a common regular 5 km by 5 km spatial grid before being used as inputs in DifERS. These re-projections are performed through simple bi-linear interpolation. The MRMS precipitation fields are remapped on the same 5 km grid through areal averaging, and temporally aggregated at the hourly resolution (with the center of each estimation hour corresponding to the time of overpass of the DMSP-F17 over the study area). The training and testing data for our study is taken during the 2021-2023 period, for which the MRMS data has been made publicly available.

\section{Model description}

\noindent Diffusion models \cite{ho2020denoising, song2021scorebased} are a recently introduced class of deep generative models which are able to sample from complex, high-dimensional distributions through an iterative denoising process. Intuitively, to generate a sample, the model starts from pure Gaussian noise and iteratively removes a small amount of noise over the course of many iterations until a realistic sample is generated. Due to their state-of-the-art performance in a wide range of domains, including images \cite{karras2022elucidating}, audio \cite{liu2023audio}, and video \cite{ma2024latte}, diffusion models have become a popular tool for generative modeling of complex data. While GANs have seen some success in precipitation nowcasting \cite{ravuri2021skilful, hayatbini2019conditional}, they are often difficult to train \cite{heusel2017gans} and can suffer from mode collapse \cite{tung2020catastrophic, bau2019seeing}, limiting their use in obtaining well-calibrated uncertainty estimates. Moreover, there is evidence that diffusion models are able to outperform GANs in perceptual quality on natural images \cite{dhariwal2024diffusion}, further motivating the use of diffusion models in precipitation retrieval. 

In what follows, we provide an informal discussion of the key ideas underpinning diffusion models. We refer to Asperti et al. \cite{asperti2023precipitation} for additional details regarding diffusion models in the context of precipitation nowcasting, and \cite{luo2022understanding} for an in-depth tutorial.

We first introduce some notation. We will use $\mathbf{x_0}$ to denote a precipitation image from our training distribution, i.e., a $64 \times 64$ pixel image representing the accumulated precipitation at 5 km resolution as observed by MRMS over a $320 \times 320$ km$^2$ subset of the study region. The variable $z$ represents the corresponding $64\times64\times D$ input vector, with $D = 23$ (10 SSMI/S channels + 13 single-band ABI images). Together, the pair $(\mathbf{x_0}, \mathbf{z})$ comprise a sample from the joint data distribution ${p_0(\mathbf{x_0}, \mathbf{z})}$. Our goal is to construct a probabilistic model ${p_\theta(\mathbf{x_0} \mid \mathbf{z})}$ of the distribution of precipitation images $\mathbf{x_0}$ conditioned on the MW/IR information contained in $\mathbf{z}$, where $\theta$ represents the set of model parameters. After learning such a model, we will be able to sample predictions $\mathbf{x_0} \sim p_\theta(\mathbf{x_0} \mid \mathbf{z})$ from this learned distribution for given covariates $\mathbf{z}$. 

\subsection{Constructing a diffusion model}

\noindent Diffusion models consist of two stochastic processes. The first, often called the \textit{forward process}, is a procedure which gradually destroys the information contained in a precipitation image $\mathbf{x_0}$ over many iterations. At the end of the forward process, the image $\mathbf{x_0}$ is turned into an image whose pixels are pure noise. The forward process is fixed prior to training the model and involves no learning. 

The second process, called the \textit{reverse process}, is obtained by running the chosen forward process backwards in time. That is, the reverse process \textit{starts} from pure noise and gradually adds more information into the image over many iterations, eventually generating an image which resembles a realistic precipitation image. However, the analytical form of the true reverse process is intractable, and thus we will need to learn an approximation of the reverse process using a neural network. 

While many versions of diffusion models have been proposed, we build upon the denoising diffusion probabilistic model (DDPM) framework of Ho et al. \cite{ho2020denoising}. To begin, we describe our choice of forward process. First, we fix a number of diffusion iterations $K$. This is typically large, and in this work we use $K = 1000$. Next, we define the forward transition densities
\begin{equation} \label{eqn:forward_distr}
    p(\mathbf{x_k} \mid \mathbf{x_{k-1}}) = \NN(\mathbf{x_k} \mid \sqrt{1 - \beta_k} \mathbf{x_{k-1}}, \beta_k \mathbf{I})
\end{equation}
for $k = 1, 2, \dots, K$. Here, $0 < \beta_k < 1$ is a fixed scalar value which controls the variance of the noise added to the image at iteration $k$. The chosen values for $\beta_1, \dots, \beta_K$ are typically referred to as the \textit{noise schedule}, and these are fixed prior to training the model. 

In essence, given an image $\mathbf{x_{k-1}}$ at iteration $k-1$, the image at the subsequent iteration $k$ is obtained via
\begin{equation} \label{eqn:forward_local}
    \mathbf{x_k} = \sqrt{1 - \beta_k} \mathbf{x_{k-1}} + \sqrt{\beta_k} \boldsymbol{\epsilon_k} \qquad \boldsymbol{\epsilon_k} \sim \NN(\boldsymbol{\epsilon_k} \mid 0, \mathbf{I}).
\end{equation}

As $0 < \beta_k < 1$, we see that this has the effect of shrinking every pixel value in $\mathbf{x_{k-1}}$ towards zero, while simultaneously adding independent Gaussian noise with variance $\beta_k$ to each pixel.

By applying the relationship in \eqref{eqn:forward_local} once again, we see that
\begin{equation*}
    \mathbf{x_k} = \sqrt{1 - \beta_k} \left( \sqrt{1 - \beta_{k-1}} \mathbf{x_{k-2}} + \sqrt{\beta_{k-1}} \boldsymbol{\epsilon_{k-1}} \right) + \sqrt{\beta_k} \boldsymbol{\epsilon_k} 
\end{equation*}
which, upon recursion, yields
\begin{equation} \label{eqn:forward_local_onestep}
    \mathbf{x_k} = \sqrt{\overline{\alpha}_k} \mathbf{x_0} + \sqrt{1 - \overline{\alpha}_k} \boldsymbol{\epsilon} \qquad \boldsymbol{\epsilon} \sim \NN(\boldsymbol{\epsilon} \mid 0, \mathbf{I})
\end{equation}
where we define the scalar constants $\alpha_k, \overline{\alpha}_k$ from the noise schedule via
\begin{equation} 
    \alpha_k = 1 - \beta_k \qquad \overline{\alpha}_k = \prod_{j=1}^k \alpha_j.
\end{equation}

In other words, the noisy image $\mathbf{x_k}$ can be obtained directly from $\mathbf{x_0}$ without having to simulate the entire forward process step-by-step. When the noise schedule $(\beta_k)_{k=1}^K$ is chosen appropriately, we will have that $\overline{\alpha}_K \approx 0$ at the terminal iteration $K$, and thus Equation \eqref{eqn:forward_local_onestep} shows us that the distribution of $\mathbf{x_K}$ is approximately $\NN(0, \mathbf{I})$. In practice, there are many viable choices for the noise schedule, and we provide additional details in the Appendix. Note that the forward process is independent of the conditioning variables $\mathbf{z}$. 

So far, we have described a procedure for iteratively turning our images into noise. To obtain a generative model, we would ideally like to know the \textit{reverse} transition densities
\begin{equation} \label{eqn:reverse_distr}
    p(\mathbf{x_{k-1}} \mid \mathbf{x_k}, \mathbf{z}) = \frac{p(\mathbf{x_k} \mid \mathbf{x_{k-1}}) p(\mathbf{x_{k-1}} \mid \mathbf{z)}}{p(\mathbf{x_k} \mid \mathbf{z})}.
\end{equation}

If we had access to these, we could start by sampling an image whose pixels are pure noise $\mathbf{x_K} \sim \NN(\mathbf{x_K} \mid 0, \mathbf{I})$, and iteratively sample $\mathbf{x_{k-1}} \sim p(\mathbf{x_{k-1}} \mid \mathbf{x_k}, \mathbf{z})$ until we obtain a draw $\mathbf{x_0} \sim p(\mathbf{x_{0}} \mid \mathbf{z})$ from our distribution of interest. Here, we now condition on the covariates $\mathbf{z}$ as we would like our model to use this information when producing a sample. However, we cannot directly compute these densities as the distributions $p(\mathbf{x_{k}} \mid \mathbf{z})$ are intractable, requiring a marginalization over the unknown $p(\mathbf{x_0} \mid \mathbf{z})$.

Thus, we require an \textit{approximation} of the true reverse densities, denoted by
\begin{equation}
    p_{\boldsymbol{\theta}}(\mathbf{x_{k-1}} \mid \mathbf{x_k}, \mathbf{z}) \approx p(\mathbf{x_{k-1}} \mid \mathbf{x_k}, \mathbf{z})
\end{equation}
where $\boldsymbol{\theta}$ represents the weights of a neural network which will be used to define this density. In practice, we choose a Gaussian distribution of the form
\begin{equation}
    p_{\boldsymbol{\theta}}(\mathbf{x_{k-1}} \mid \mathbf{x_k}, \mathbf{z}) = \NN(\mathbf{x_{k-1}} \mid \mu_{\boldsymbol{\theta}}(k, \mathbf{x_k}, \mathbf{z}), \beta_k \mathbf{I}).
\end{equation}

The form of this distribution is a heuristic which has been found to work well in practice \cite{ho2020denoising}. Here, a neural network $\mu_{\boldsymbol{\theta}}(k, \mathbf{x_k}, \mathbf{z})$ takes in the iteration $k$, the noisy image $\mathbf{x_k}$, and the conditioning information $\mathbf{z}$, and is tasked with predicting the expected value of $\mathbf{x_{k-1}}$ given this information. In this way, our neural network may be viewed as performing a denoising operation.

To train the model, we would like to minimize the model's negative log-likelihood $\E \left[- \log p_{\boldsymbol{\theta}}(\mathbf{x_0} \mid \mathbf{z}) \right]$, where this expectation is taken over samples $(\mathbf{x_0}, \mathbf{z}) \sim p_0(\mathbf{x_0}, \mathbf{z})$ from the data distribution. However, evaluating $p_{\boldsymbol{\theta}}(\mathbf{x_0} \mid \mathbf{z})$ is intractable as this requires marginalizing over the intermediate diffusion steps $\mathbf{x_1}, \dots, \mathbf{x_K}$. Instead, we take a variational approach, and minimize an upper bound on this likelihood. Although deriving this variational upper bound is nontrivial, it can be shown \cite{ho2020denoising, salimans2022progressive} that it is given by
\begin{multline}
    \E \left[ w(k) \norm{\frac{1}{\sqrt{\alpha_k}}\left(\mathbf{x_k} - \frac{\beta_k}{\sqrt{1 - \overline{\alpha}_k}} \boldsymbol{\epsilon} \right) - \mu_{\boldsymbol{\theta}}(k, \mathbf{x_k}, \mathbf{z)}}^2 \right]
\end{multline}
where this expectation is taken over $(\mathbf{x_0}, \mathbf{z}) \sim p_0(\mathbf{x_0}, \mathbf{z})$, ${\boldsymbol{\epsilon} \sim \NN(0, \mathbf{I})}$ and $k \sim \text{Unif}(\{1, 2, \dots, K \})$. Here, $w(k)$ is a scalar which corresponds to a different weighting of the loss at each diffusion step. Observe that this loss is merely a mean-squared error, where the target involves the noisy image $\mathbf{x_k}$, the noise $\boldsymbol{\epsilon}$, and constants arising from the noise schedule. Given that our target has this known structure, it is natural to parametrize the model's output $\mu_{\boldsymbol{\theta}}(k, \mathbf{x_k}, \mathbf{z})$ as
\begin{equation} \label{eqn:eps_to_mu}
    \mu_{\boldsymbol{\theta}}(k, \mathbf{x_k}, \mathbf{z}) = \frac{1}{\sqrt{\alpha_k}} \left(\mathbf{x_k} - \frac{\beta_k}{\sqrt{1 - \overline{\alpha}_k}} \epsilon_{\boldsymbol{\theta}}(k, \mathbf{x_k}, \mathbf{z}) \right)
\end{equation}

where $\epsilon_{\boldsymbol{\theta}}(k, \mathbf{x_k}, \mathbf{z})$ is a neural network trained not to predict the less-noisy $\mathbf{x_{k-1}}$, but rather to directly predicts the noise $\boldsymbol{\epsilon}$ added to $\mathbf{x_{k-1}}$ to produce $\mathbf{x_k}$. Note that this neural network produces an image having the same dimensions as the input $\mathbf{x_k}$. This parametrization has the benefit of incorporating known information from the forward process into the model parametrization -- in other words, since the model has direct access to $\mathbf{x_k}$, we may use this to directly parametrize the mean $\mu_{\boldsymbol{\theta}}(k, \mathbf{x_k}, \mathbf{z})$ and only use the neural network to predict the necessary missing information. 

Under this parametrization, we may train the model by minimizing the simplified loss \cite{ho2020denoising}, given by
\begin{equation} \label{eqn:eps_loss}
    L(\boldsymbol{\theta}) = \E_{\mathbf{x_0}, \mathbf{z}, \boldsymbol{\epsilon}, k} \left[ \lvert\lvert \boldsymbol{\epsilon} - \epsilon_{\boldsymbol{\theta}}(k, \mathbf{x_k}, \mathbf{z}) \rvert\rvert^2 \right].
\end{equation}

Note that we have set $w(k) = 1$, a standard heuristic in diffusion modeling, as this weighting has been observed to produce higher-quality samples. We note that other model parametrizations are possible, and we discuss the details in the Appendix. The expectation in the loss is estimated by Monte Carlo sampling, allowing us to easily compute the gradient of $L(\boldsymbol{\theta})$ via backpropagation for training.

\subsection{Model architecture}
\begin{figure*}
    \centering
    \includegraphics[width=.8\linewidth]{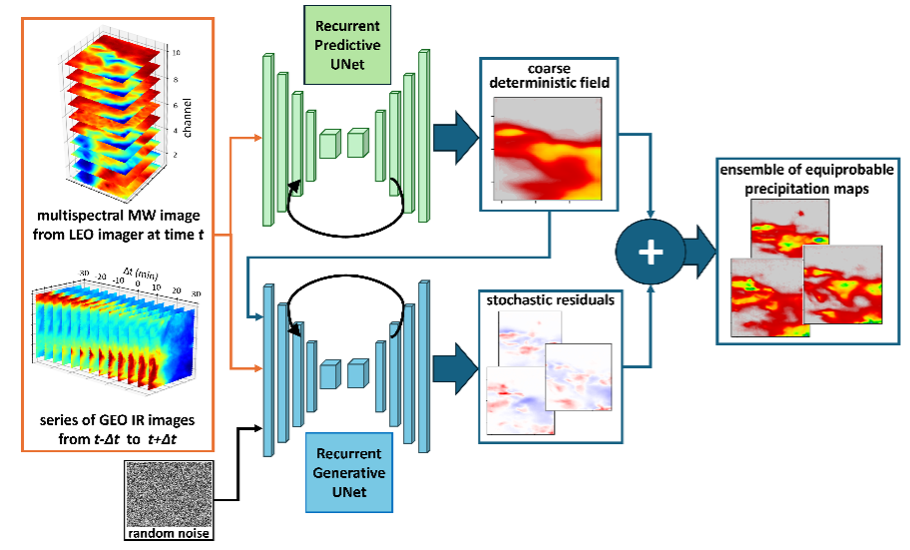}
    \caption{Schematic representation of the DifERS architecture. The inputs are 64×64-pixel (320 km × 320 km) brightness temperature fields, specifically, 10 fields corresponding to the 10 channels of SSMI/S at time t, and 13 fields corresponding to ABI brightness temperature at 10.3 µm at times (t-30min, t-25min, t-20min, …, t+30min). The outputs are 64×64-pixel hourly precipitation maps (precipitation height accumulated from t-30min to t+30min). For each observed scene, 128 precipitation maps are sampled from the model.}
    \label{fig:2}
\end{figure*}

\noindent A key choice in obtaining high-quality results with diffusion models is the choice of a model architecture for $\epsilon_{\boldsymbol{\theta}}(k, \mathbf{x_k}, \mathbf{z})$ that is appropriate for modeling the spatial characteristics of the problem at hand. In this work, we propose a two-stage architecture, consisting of two distinct UNets \cite{ronneberger2015u}. The UNet architecture is a type of convolutional neural network which takes in an image (plus additional covariates) and produces a new image having the same dimensions as the original input, which is necessary for implementing a diffusion model as described. This output image is produced by a series of convolutions and rescaling operations, allowing the model to incorporate information across various spatial scales. 

In our model, the first UNet seeks to produce a coarse deterministic field, which can be viewed intuitively as an estimate of the conditional mean of $\mathbf{x_0}$ given $\mathbf{x_k}$ and $\mathbf{z}$. However, simply predicting the conditional mean can result in blurry, overly smooth images, as is known to be a common failure case of MSE-based approaches \cite{Guilloteau2023Constraining,zhao2017towards}. Thus, the second UNet takes the output of the first UNet (along with the noisy image $\mathbf{x_k}$ and conditioning variables $\mathbf{z}$) and predicts a stochastic residual which seeks to add additional fine-grained details. The generative process of our architecture is visualized in Figure \ref{fig:2}. We note that similar residual diffusion architectures have shown strong performance on precipitation downscaling tasks \cite{srivastava2024precip, mardani2023generative}, further motivating our approach. 

The model is trained using Adam \cite{kingma2014adam}, a standard stochastic gradient descent based method, on a set of 3617 scenes sampled at the times SSMI/S overpasses over the study domain. Our training data is normalized such that all pixel values lie in the range $[-1, 1]$. We provide additional details on our architecture and training procedure in the Appendix.

\section{Results}

\subsection{Model accuracy and performance at multiple spatial scales}

\noindent In this section we compare the hourly precipitation fields generated by DifERS to the MRMS reference fields. The statistical resemblance of the individual ensemble members to the MRMS fields is verified in terms of the distribution of precipitation intensities, and of the multiscale variance and spatial autocovariance through the Fourier power spectrum. The ensemble mean is considered to be equivalent to a Bayesian posterior mean estimate and is evaluated against the MRMS “truth” in terms of mean squared error (MSE) and coefficient of linear correlation. The Fourier spectral coherence between the ensemble members and the MRMS truth is also analyzed to evaluate the ability of DifERS to properly localize precipitation features. The evaluation dataset, disjoint from the training dataset, comprises 486 precipitation scenes of dimensions 64×64 pixels at 5 km resolution.

Coincident precipitation fields from two operational satellite products, the MW-based SSMI/S-GPROF V7 product  \cite{Randel2020GPROF} and the IR-based PERSIANN-CCS product \cite{Hong2004PERSIANN} are also evaluated against MRMS over the same set of precipitation scenes. The PERSIANN-CSS and SSMI/S-GPROF fields are both reprojected on the 5 km grid of the DifERS outputs for the evaluation, their original nominal resolutions being respectively 4 km and 38 km. The temporal resolution of PERSIANN-CCS is one hour. However, it should be noted that the PERSIANN-CCS estimates are temporal integrals corresponding to calendar hours (from X:00 GMT to X+1:00 GMT), and therefore don’t perfectly match the one-hour period centered on the time $t$ of the SSMI/S overpass. GPROF estimates are per-design estimates of the instantaneous precipitation intensity at time $t$; they are however generally considered to be fairly representative of the 20-min-to-1-h integrated precipitation intensity \cite{kidd2021theglobal,Turk2009validating}. Including PERSIANN-CSS and SSMI/S-GPROF in the evaluation allows to define baseline targets for the expected accuracy of DifERS. It should be noted that, while in its present setup DifERS is trained to reproduce MRMS precipitation maps in southeastern US specifically, PERSIANN-CSS and GPROF are optimized for global performance. 

\begin{figure*}
    \centering
    \includegraphics[width=.85\linewidth]{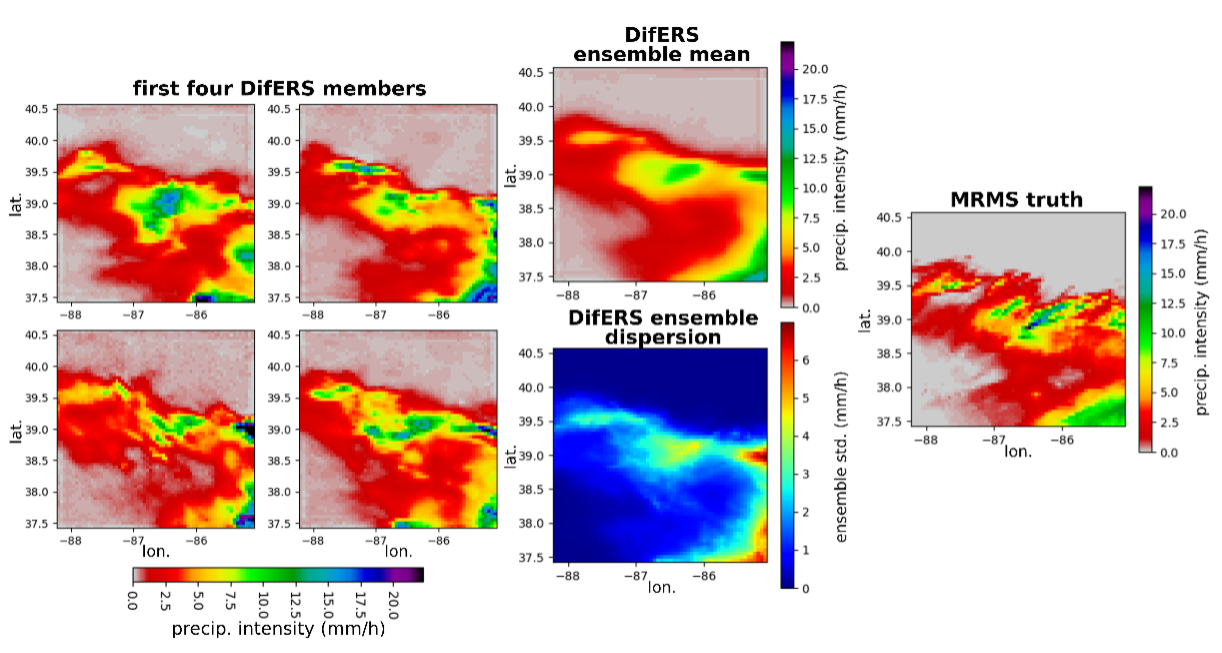}
    \caption{Precipitation fields generated by DifERS for the period ranging from 12:40 UTC to 13:40 UTC on 2021-05-04. Four of the 128 generated ensemble members are shown, along with the ensemble mean, ensemble dispersion (standard deviation) and the MRMS ground truth. }
    \label{fig:3}
\end{figure*}
\begin{figure*}
    \centering
    \includegraphics[width=.85\linewidth]{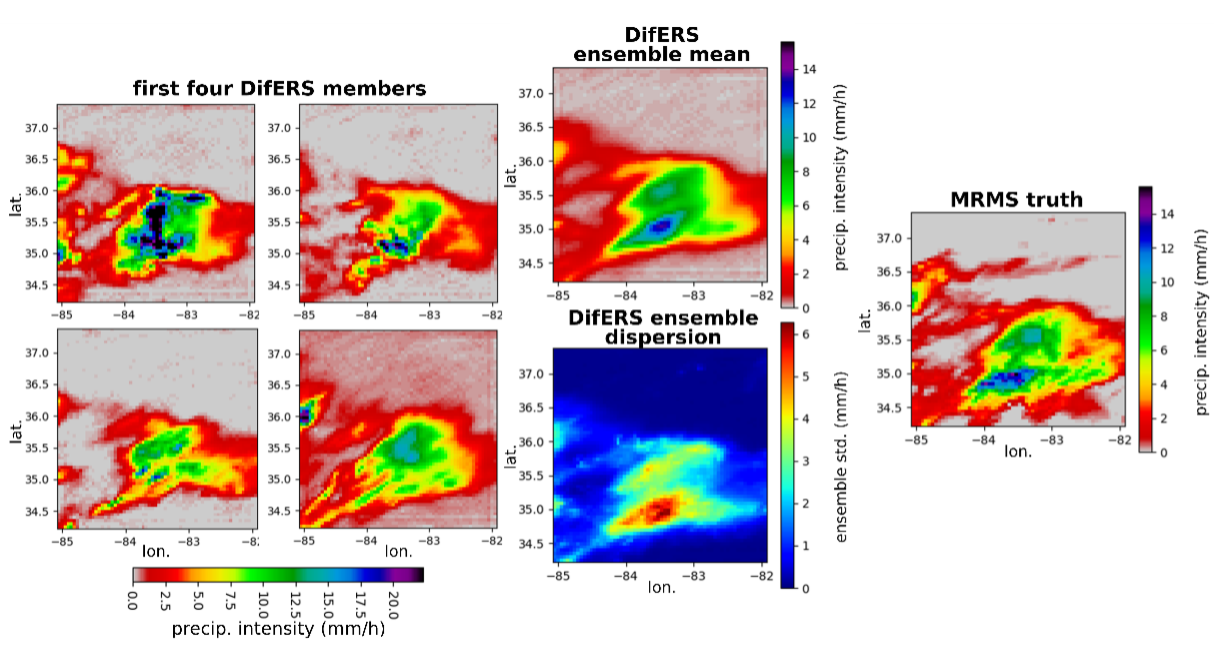}
    \caption{Same as Figure 3, over a different portion of the study domain, for the period ranging from 12:30 UTC to 13:30 UTC on 2020-10-28. }
    \label{fig:4}
\end{figure*}

We first qualitatively evaluate the visual resemblance between the DifERS and MRMS precipitation fields. Figures 3 and 4 show the first four members of the DifERS ensemble for two randomly-selected scenes of the evaluation dataset; the ensemble mean and ensemble dispersion (across all 128 ensemble members) are also shown along with the MRMS “ground truth”. For both scenes, the ensemble members are visually similar to the MRMS truth in terms of the approximate location and magnitude of the precipitation features, and in terms of the texture of the precipitation fields. The ensemble mean is naturally smoother than the individual ensemble members, with maximal values around 13 and 12 mm/h in the figures 3 and 4 cases respectively, against 23 and 16 mm/h maximal values in the individual ensemble members or in the MRMS fields.
\begin{figure}
    \centering
    \includegraphics[width=0.8\linewidth]{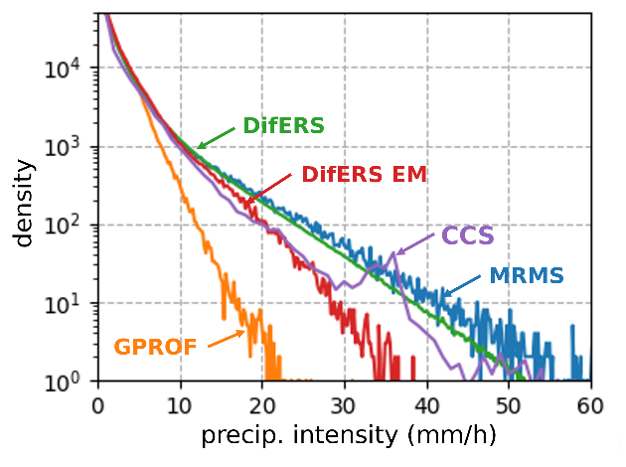}
    \caption{Density histograms showing the statistical distribution of precipitation intensities at 5 km and one hour resolution in the DifERS fields, along with the distributions for the DifERS ensemble mean (EM) field, for the MRMS ground truth and for the operational GPROF and PERSIANN-CCS products.}
    \label{fig:5}
\end{figure}
\begin{table}[t]
\caption{(left column) KLD of the statistical distribution of pixel precipitation intensities at the 5 km resolution to the MRMS distribution. (right column) KLD of the Fourier power spectrum of the precipitation maps to the MRMS power spectrum.}
    \centering
    \begin{tabular}{lcc}
    \toprule
    & Distribution KLD & Fourier Spectrum KLD \\
    \midrule
    DifERS & $2 \times 10^{-4}$ & $11 \times 10^{-4}$ \\ 
    DifERS EM & $35 \times 10^{-4}$ & $104 \times 10^{-4}$ \\
    CCS & $79 \times 10^{-4}$ & $119 \times 10^{-4}$ \\
    GPROF & $165 \times 10^{-4}$ & $299 \times 10^{-4}$ \\
    \end{tabular}
    \label{tab:1}
    
\end{table}

Across the whole evaluation dataset, the statistical distribution of pixel precipitation intensities in the DifERS fields closely matches that of the MRMS fields (Figure 5), with only a slight underestimation of the frequency of occurrence of precipitation rates above 30 mm/h. In comparison, for the GPROF statistical distribution, the occurrence of all precipitation intensities above 7 mm/h is dramatically underestimated. However, it should be recalled that, while the GPROF fields are projected on a 5 km grid in this study, the original resolution of SSMI/S-GPROF is 38 km. For the PERSIANN-CCS distribution, the occurrence of precipitation intensities between 10 and 33 mm/h is significantly underestimated, as well as that of precipitation rates above 35 mm/h. Notably, the distribution of precipitation intensities in PERSIANN-CCS shows a peak at 35 mm/h. Regarding the DifERS ensemble mean, its smoothness is apparent in the statistical distribution, as intensities above 15 mm/h are “compressed” when the different members of the ensemble are averaged together. In addition to the visual evaluation of the distributions, to provide a quantitative measure of the ability of DifERS to reproduce the statistical distribution of MRMS precipitation intensities, we compute the Kullback-Leibler divergence (KLD) between the DifERS and MRMS discretized distributions (see Appendix A for the formal definition of the KLD), using a 1 mm/h discretization step over the 0 to 52 mm/h intensity range. The lower the KLD, the more similar the two disctributions. The computed KLD of the DifERS distribution to the MRMS distribution is $2\times10^{-4}$, which is several orders of magnitude lower than the values computed for the GPROF and PERSIANN-CCS products, reported in Table 1. As expected, the KLD of the DifERS ensemble mean distribution to the MRMS distribution is much higher than for the individual members, at $35 \times 10^{-4}$.
\begin{figure*}
    \centering
    \includegraphics[width=0.9\linewidth]{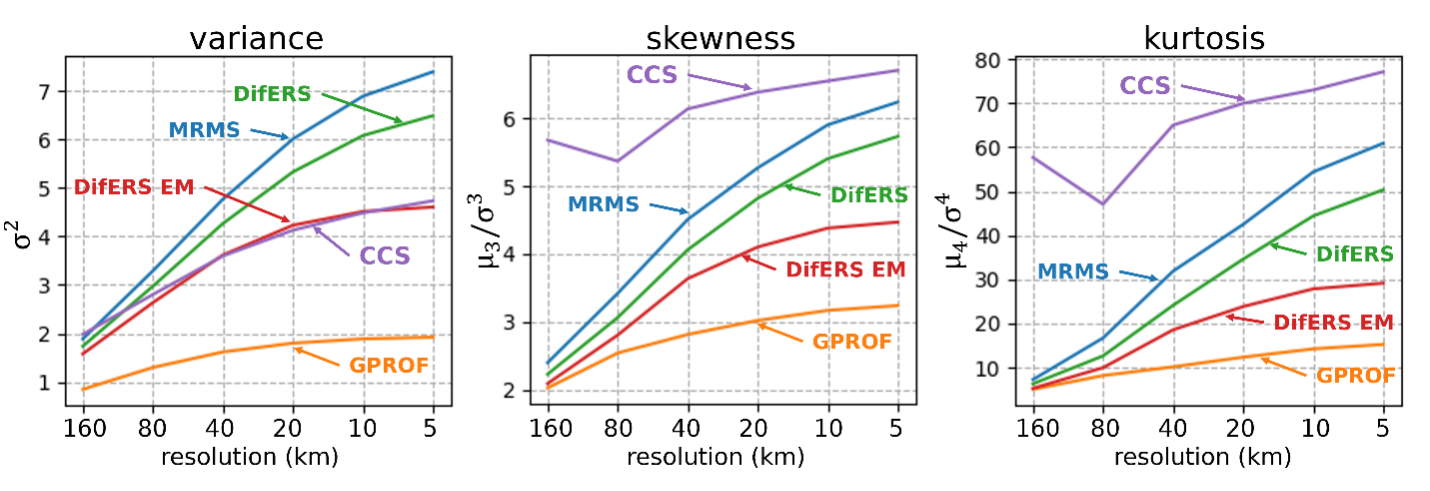}
    \caption{Variance, skewness and kurtosis coefficients of the statistical distribution of hourly precipitation intensities as a function of the spatial resolution, for the DifERS fields, the DifERS ensemble mean (EM) fields, the GPROF fields, the PERSIANN-CCS fields and the MRMS ground truth.}
    \label{fig:6}
\end{figure*}

To verify that DifERS can reproduce the statistical distribution of precipitation intensities across multiple spatial scales, and not only at the 5 km pixel scale, we look at the second, third and fourth moments of the distributions (variance, skewness and kurtosis) at different resolutions between 5 and 160 km (Figure 6). The manner in which all three moments vary with scale is consistent between DifERS and MRMS. At resolutions finer than 40 km, the variance, skewness coefficient and kurtosis coefficient of the DifERS ensemble mean are significantly lower than for the individual ensemble members, once again illustrating the smoothing effect of the ensemble averaging, leading to “compressed” statistical distributions. Another way to assess the multiscale properties of the precipitation maps is to analyze their Fourier power spectra, which are directly related to the spatial autocorrelation of the fields \cite{Cohen1998WK}. Figure 7 shows that the DifERS spatial power spectrum is highly consistent with the MRMS spectrum. As for the statistical distributions the (dis)similarity between two power spectra can be measured through the KLD, given that power spectra are formally equivalent to distributions. The KLD of the power spectra of the different satellite precipitation estimates to the MRMS power spectrum are reported in Table 1.
\begin{figure}
    \centering
    \includegraphics[width=.8\linewidth]{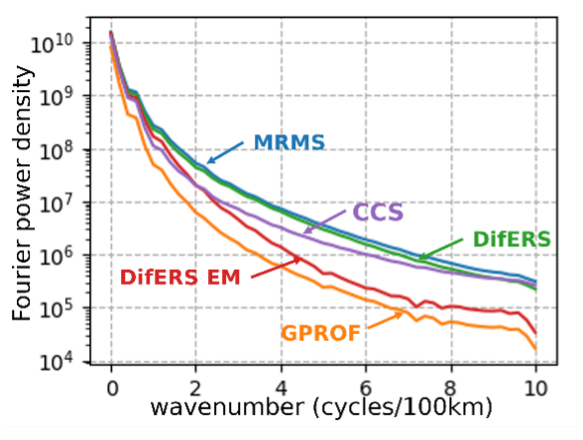}
    \caption{Omnidirectional spatial Fourier power spectrum of the DifERS precipitation intensity maps at 5 km and one hour resolution, along with the spectra of the DifERS ensemble mean (EM), of the MRMS ground truth and of the operational GPROF and PERSIANN-CCS products.}
    \label{fig:7}
\end{figure}
\begin{figure}
    \centering
    \includegraphics[width=0.8\linewidth]{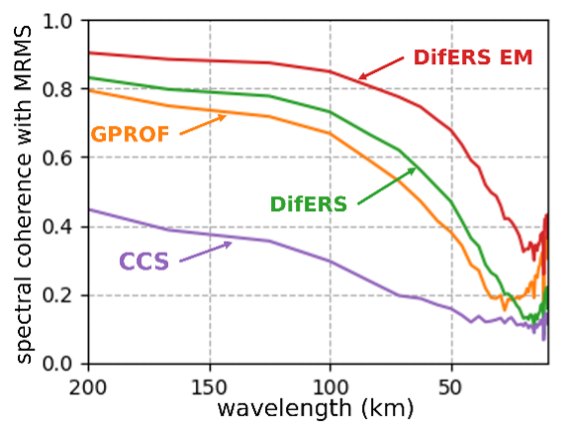}
    \caption{Omnidirectional spatial Fourier coherence between DifERS precipitation intensity maps at 5 km and the MRMS ground truth. The coherence with MRMS is also shown for the DifERS ensemble mean (EM) fields, the GPROF fields and the PERSIANN-CCS fields.}
    \label{fig:8}
\end{figure}
\begin{table}
\caption{Performance metrics of the DifERS ensemble mean, PERSIANN-CSS and GPROF when compared with the MRMS ground truth at 5 km resolution. The metrics are the coefficient of linear correlation (CC), the mean squared error (MSE), the relative mean bias, and the effective resolution (ER). }
\centering
\label{tab:2}
\begin{tabular}{lcccc}
\toprule
 & CC & MSE (mm$^2$h$^{-2}$) & Mean Bias & ER (km) \\
\midrule
DifERS EM & 0.82 & 2.48 & +2\% & 23 \\
CCS & 0.35 & 8.04 & -29\% & $>$100 \\
GPROF & 0.67 & 4.31 & -26\% & 60 \\
\end{tabular}
\end{table}

In addition to verifying that the DifERS fields look realistic and have adequate multiscale statistical properties, we verify that DifERS properly localizes the precipitation features. For this, we compute the Fourier spectral coherence between the DifERS fields (individual realizations and ensemble mean) and the MRMS fields (Figure 8). In terms of the coherence with MRMS, the DifERS ensemble mean shows higher values than the individual ensemble members, which is expected as the stochastically-generated spatial variability in the individual members inevitably reduces the coherence. For the ensemble mean, the coherence with MRMS is higher than 0.7 down to the 56 km wavelength. DifERS is therefore capable of accurately resolving the spatial variability of precipitation down to the 23 km scale (half of the shortest resolved wavelength in accordance with the Nyquist–Shannon theorem), which is approximately the instrumental resolution of SSMI/S. This defines what we call the effective resolution of DifERS (the 0.7 spectral coherence threshold corresponds to a 1:1  spectral signal-to-noise ratio \cite{Guilloteau2021howwell}). The integral of the spectral coherence across all wavenumbers determines the coefficient of linear correlation between DifERS and MRMS. The coefficient of linear correlation is a widely used metric to assess the overall consistency between two different estimates of a variable in geophysical remote sensing; another widely used metric is the mean squared difference (MSD). These metrics for the DifERS ensemble mean (and for GPROF and PERSIANN-CCS) against MRMS are reported in Table 2, along with the mean bias and the effective resolution. For all metrics, the DifERS ensemble mean shows excellent performance, well above the GPROF and PERSIANN-CCS baseline.

\subsection{Probabilistic interpretation of the outputs and uncertainty quantification}
\begin{figure}
    \centering
    \includegraphics[width=0.8\linewidth]{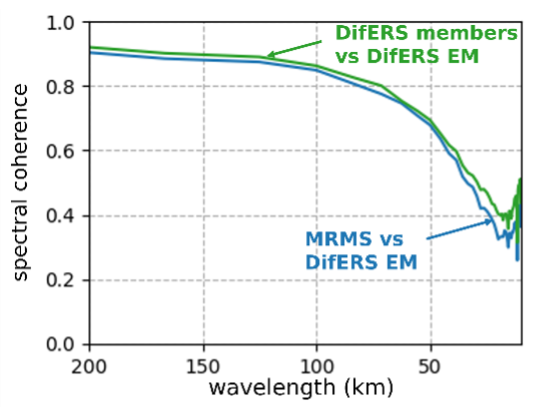}
    \caption{Spatial Fourier coherence between the DifERS ensemble members and the DifERS ensemble mean, compared to the coherence between the MRMS and the DifERS ensemble mean.}
    \label{fig:enter-label}
\end{figure}
\begin{figure*}
    \centering
    \includegraphics[width=0.7\linewidth]{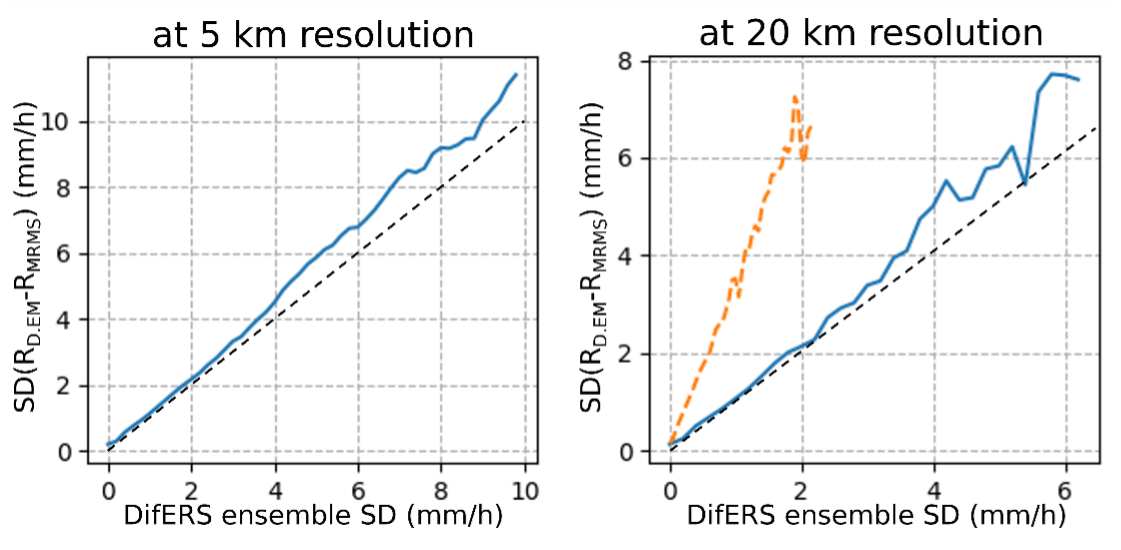}
    \caption{(left, solid blue curve) Empirical standard deviation of the retrieval error $R_{D.EM}-R_{MRMS}$ as a function of the standard deviation of the 128 DifERS ensemble in 5 km pixels. (right, solid blue curve). Empirical standard deviation of the retrieval error $R_{D.EM}-R_{MRMS}$ as a function of the standard deviation of the 128 DifERS ensemble in 20 km area-averaged pixels. To produce these curves, the pixels of the evaluation dataset are grouped into 50 bins according to standard deviation of the 128 DifERS ensemble, and the standard deviation of the retrieval error $R_{D.EM}-R_{MRMS}$ is then computed in each bin. Each bin contains at least 100 pixels. The orange dashed curve in the right panel is what we obtain if we ignore the spatial correlation of the scholastic variability in the ensemble members and compute the ensemble variance at 20 km resolution (4×4 aggregated pixels) as the mean of the 5-km pixel variances divided by $4^2$. }
    \label{fig:enter-label}
\end{figure*}

\noindent In this section we show how the dispersion of the DifERS ensemble can provide a measure of the estimation uncertainty at any desired spatial scale. We also demonstrate how the ensemble can be interpreted as a probabilistic estimation. In the previous section we showed that the individual members of the DifERS ensemble are similar to the MRMS truth in terms of their multiscale statistical properties. Therefore, if we were to “hide” the MRMS truth among the DifERS ensemble members, it would not stand out for being statistically different. Figure 9 shows the spatial Fourier coherence spectrum between the DifERS ensemble members and the DifERS ensemble mean, along with the coherence spectrum between the MRMS and the DifERS ensemble mean. The two coherence spectra are remarkably similar, being quasi on top of each other. This shows that, on average, the coherence between any ensemble member and the ensemble mean is essentially the same as the coherence between the ensemble mean and the MRMS truth. Once again, the MRMS truth would not stand out if put among the DifERS ensemble members; on average, in terms of their relative coherence, any ensemble member is as similar to any other ensemble mean as is to the MRMS truth. This demonstrates that the dispersion of the ensemble members can be used as a measure of the retrieval uncertainty. 

In figures 3 and 4 the DifERS ensemble dispersion was represented as the standard deviation of the 128 ensemble values in each 5 km pixel. To evaluate the meaningfulness of this standard deviation measure for uncertainty quantification, we group all the pixels of the evaluation datasets according to their DifERS ensemble standard deviation into 50 bins between 0 and 10 mm/h. For each bin we then compute the empirical standard deviation of the true retrieval error $R_{D.EM} - R_{MRMS}$ (where $R_{D.EM}$ is the DifERS ensemble mean value and $R_{MRMS}$ is the MRMS value). As shown in Figure 10 (left), the resulting curve remains close to the 1:1 line, demonstrating that, for each bin, the standard deviation of the error $R_{D.EM} - R_{MRMS}$ is approximately equal to the DifERS ensemble standard deviation value used to define the bin. Only for values higher than 4 mm/h, the true standard deviation of the error is slightly larger on average than the ensemble standard deviation. This small difference between the true standard deviation of the error and the ensemble standard deviation can be attributed to the epistemic uncertainty, as the DifERS ensemble generation only accounts for the aleatoric uncertainty. 

The evaluation of the ensemble standard deviation is repeated after coarsening the DifERS and MRMS fields at the 20 km resolution through simple pixel area averaging (Figure 10, right). At the 20 km resolution also, the ensemble standard deviation is found to be a good proxy of the retrieval error standard deviation. This is because the stochastically-generated variability in each DifERS ensemble member is spatially coherent, which allows us to account for the spatial correlation of retrieval errors, and to properly estimate the retrieval uncertainty at any desired resolution between 5 and 320 km. This would not be possible if the stochastic variability was generated in each pixel independently. If we were to ignore the spatial correlation of retrieval errors and estimate the error variance in aggregated pixels as the sum of the 5-km pixel variances divided by the number of averaged pixels squared, we would dramatically underestimate the uncertainty at coarsened resolutions, as shown in Figure 10 (right, orange dashed curve).      
Beyond the standard deviation of the retrieval error, interpreting the DifERS ensemble as a probabilistic estimation, we can derive the probability of exceedance of a given intensity value for any pixel, simply by computing the fraction of the 128 ensemble members with intensity above the chosen value in this pixel. This can be done at the 5 km pixel resolution, or at any desired coarsened resolution. Figure 11 shows the probability of exceeding the 1 mm/h, 4 mm/h and 10 mm/h intensity thresholds derived from the DifERS ensemble in each 5 km pixel for the scene shown in Figure 3, (12:40 UTC to 13:40 UTC on 2021-05-04). As we did with the ensemble standard deviation, we regrouped the pixels of the evaluation dataset in 50 bins according to the DifERS-derived probability of exceeding the 1 mm/h intensity threshold; we then computed for each bin the actual fraction of the corresponding MRMS pixels with intensity above the threshold. This was repeated for the 2 mm/h, 4 mm/h and 8 mm/h thresholds; the results are shown in Figure 12 (top row). For all four tested thresholds, the curve showing the MRMS fraction above the threshold against the DifERS ensemble probability of exceedance closely follows the 1:1 line. This demonstrates that the DifERS-derived probability of exceedance matches the MRMS exceedance rate on average over the evaluation dataset. This was also verified at the 20 km resolution (Figure 12, bottom row), with satisfactory results, the increased noisiness in the curves at coarser resolution and higher thresholds is only due to the smaller sample size (number of pixels per bin) available for computing the MRMS exceedance rate.
\begin{figure*}
    \centering
    \includegraphics[width=0.8\linewidth]{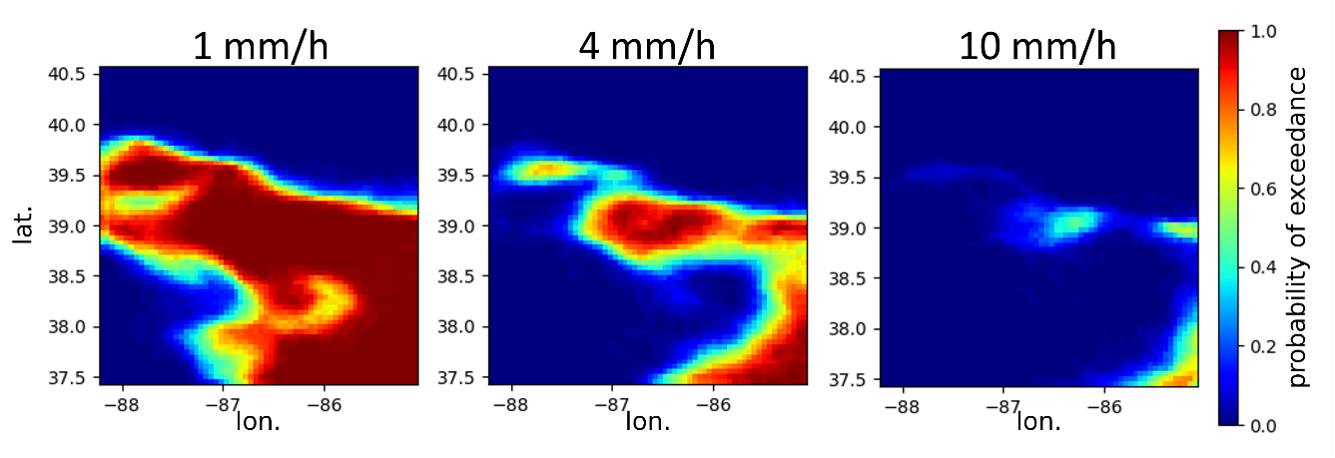}
    \caption{Probability of exceedance of the 1 mm/h, 4 mm/h and 10 mm/h intensity thresholds derived from the DifERS ensemble for the scene shown in Figure 3, (12:40 UTC to 13:40 UTC on 2021-05-04).  }
    \label{fig:11}
\end{figure*}
\begin{figure*}
    \centering
    \includegraphics[width=0.8\linewidth]{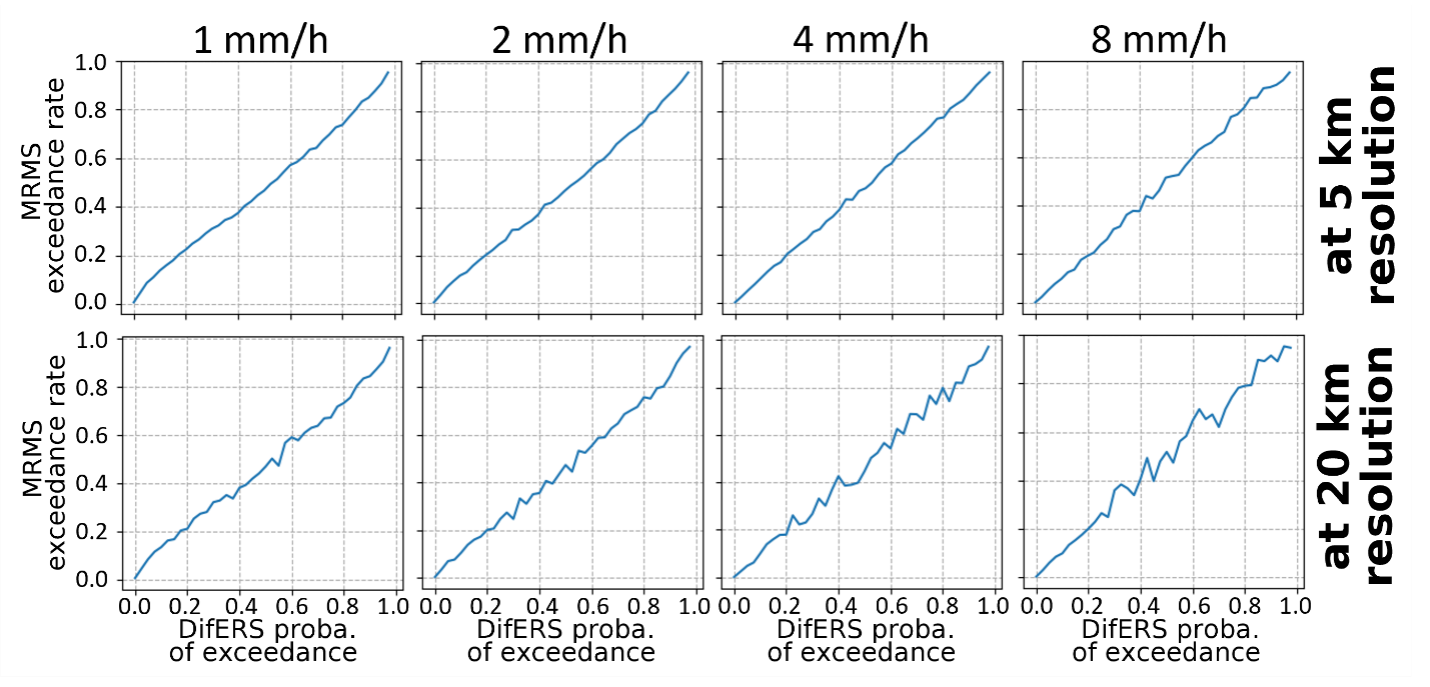}
    \caption{Average MRMS exceedance rate at 1, 2, 4 and 8 mm/h as a function of the pixel DifERS-derived probability of exceedance at 5-km resolution (top) and at 20-km resolution (bottom). To produce these curves, the pixels of the evaluation dataset are grouped into 50 bins between 0 (0\%) and 1 (100\%), according to the DifERS-derived probability of exceedance, and the corresponding MRMS exceedance rate is then computed in each bin.}
    \label{fig:enter-label}
\end{figure*}
\begin{figure*}
    \centering
    \includegraphics[width=0.8\linewidth]{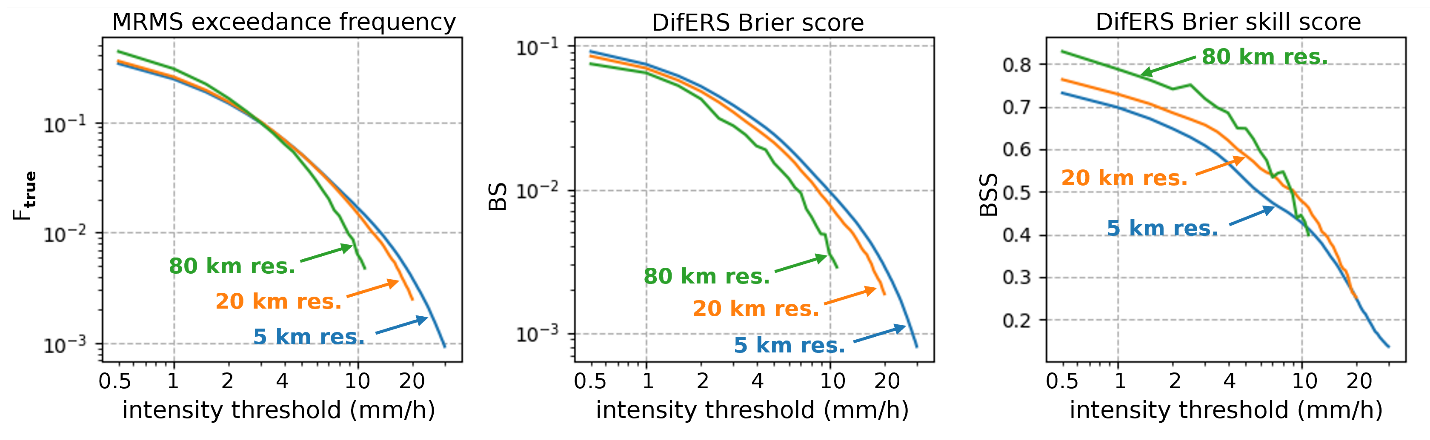}
    \caption{(left) MRMS exceedance frequency $F_\text{true}$ of precipitation intensity $R_t$ as a function of $R_t$, at 5 km, 20 km and 80 km resolutions. (center) Brier score BS of the DifERS-estimated probability of exceedance of precipitation intensity $R_t$, as a function of $R_t$, at 5 km, 20 km and 80 km resolutions. (right) Brier skill score $BSS=1-BS/F_\text{true}$, of the DifERS-estimated probability of exceedance of precipitation intensity $R_t$, as a function of $R_t$, at 5 km, 20 km and 80 km resolutions. All statistics are computed over the 486 scenes of the evaluation dataset.}
    \label{fig:enter-label}
\end{figure*}

To complete the evaluation of the DifERS ensemble as a probabilistic estimate we utilize the Brier score. The Brier score is a strictly proper scoring rule for evaluating probabilistic predictions of the occurrence of an event \cite{Brier1950}. Here, the event is the exceedance of a given intensity threshold in a given pixel. The Brier score is computed as an average over a set of multiple predictions as the mean squared difference between the binary truth (1 if the event occurs and 0 if it doesn’t occur), and the probabilistic prediction (continuous value between 0 and 1). See Appendix A for the formal definition of the Brier score. The lower the Brier score, the more skillful the prediction. The Brier score BS of the DifERS-predicted probability of exceedance at values ranging from 0.5 mm/h to 30 mm/h is computed here, at the 5 km, 20 km and 80 km resolutions (Figure 13). One can see that, at all resolutions, BS is lower at higher intensity thresholds. This is naturally expected, as the true frequency of exceedance $F_{\text{true}}$ decreases when the threshold increases. For a skillful predictor, BS shall be always lower than $F_{\text{true}}$. Indeed, with a “no-skill” probabilistic prediction systematically predicting a 0\% probability of occurrence, BS would be equal to $F_{\text{true}}$. The right panel of Figure 13 shows the Brier skill score $BSS = 1 - BS/F_{\text{true}}$, as a normalized measure of the skill of the DifERS-estimated probability of exceedance. The best achievable  value is 1, and a 0 value (or a negative value) indicates a no-skill prediction. At all resolutions the BSS decreases when the intensity threshold increases, revealing that the occurrence of high-intensity precipitation is more challenging to estimate from passive satellite measurements than that of low- and medium-intensity precipitation.  The BSS however remains positive for all intensity thresholds up to 30 mm/h and shows values higher than 0.5 for intensity thresholds between 0.5 mm/h and 6 mm/h. The probabilistic estimation skill of DifERS increases when aggregating the outputs at coarser resolutions, as the BSS is higher at the 20 km and 80 km resolutions than at the 5 km resolution (at least up to the 6 mm/h intensity threshold). This attests that the ensemble dispersion decreasing at coarser scales, as shown earlier (Figure 6), truly reflects an increase of the estimation accuracy with spatial coarsening.

\section{Conclusion and perspectives}

\noindent The results of the present article demonstrate the utility of deep neural diffusion models for generating probabilistic ensembles of precipitation maps conditioned on multisensor satellite passive radiometric measurements in the MW and IR domain. The experimental DifERS model was trained to generate hourly precipitation maps at 5 km resolution over the southeastern US conditioned on SSMI/S and ABI observations. Not only is the DifERS model capable of generating realistic-looking precipitation fields, consistent with the “ground truth” fields derived from the MRMS gauge-radar network in terms of their multiscale statistical properties, but it is also capable of utilizing the satellite information to generate precipitation features with location and magnitude coinciding with that of the ground truth.        

The DifERS model was set to generate a 128-member ensemble of possible precipitation maps for each observed scene. This ensemble is interpreted as a probabilistic estimation, with each ensemble member being an equiprobable realization. Under this assumptions the ensemble mean thus corresponds to the Bayesian posterior mean, which is a minimum MSE estimator. Over the validation dataset the DifERS ensemble mean shows high consistency with the MRMS ground truth, with a 0.82 linear correlation between the two. The coherence between the MRMS truth and any member of the DifERS ensemble is found to be nearly the same as the coherence between any two DifERS ensemble members on average over the validation dataset. The statistical dispersion of the DifERS ensemble members thus provides a measure of the estimation uncertainty, at the original 5 km pixel scale or at any desired aggregated scale. The probability of exceedance of any given precipitation intensity threshold can also be derived from the DifERS ensemble, at any location and any desired spatial resolution.
	
 In addition to the utility of their stochastic generative capabilities for probabilistic ensemble generation, deep neural diffusion models allow one to easily combine information from different sources as observational constraints. Here IR and MW observations were combined together to condition the generation of the precipitation maps. It is worth noting that, while DifERS produces precipitation maps at the 5-km resolution, the footprint size of SSMI/S is of the order of 40 km. The DifERS scheme can therefore be seen as both a retrieval scheme and a downscaling scheme, utilizing the high-resolution IR information. A sensitivity analysis (not shown) revealed that the multispectral MW information from SSMI/S provides on average a stronger constraint than the IR information from ABI in DifERS. Currently, the available constellation of passive MW sensors (including all SSMI/S instruments and other instruments with similar capabilities \cite{kidd2021theglobal}) allows for 8 to 12 overpasses per day at any point of the globe. While IR-only algorithms like PERSIANN-CCS can provide estimations every one hour (or even less) at any point of the globe (excluding latitudes above 65° where geometrical distortions make geostationary imagery hardly usable), precipitation estimates relying on MW sensors need to be “propagated” or dynamically interpolated through space and time to produce continuous estimates through the day \cite{huffman2020IMERG,Ushio2009GSMAPmvk,Joyce2011kalman}. To fully exploit ABI 5-min temporal resolution, an architecture similar to that of DifERS could be used to temporally “interpolate” precipitation maps at 5 minutes between two overpasses of a MW sensor, and using the IR images as a constraint to guide the interpolation. 

\section*{Acknowledgements}

\noindent The authors acknowledge the support of NASA through the Precipitation Measurement Mission program (grant 80NSSC22K0597) and the Weather and Atmospheric Dynamics program (grant 80NSSC23K1304), as well as the support of NSF's Division of Information and Intelligent Systems through the Expand AI2ES program (grant IIS2324008). This research was supported
in part by the Hasso Plattner Institute (HPI) Research Center in Machine Learning and Data Science at the University of California, Irvine, and by the US National Science Foundation under award 1900644.












\bibliographystyle{IEEEtran}
\bibliography{refs}


\begin{IEEEbiography}[{\includegraphics[width=1in,height=1.25in,clip,keepaspectratio]{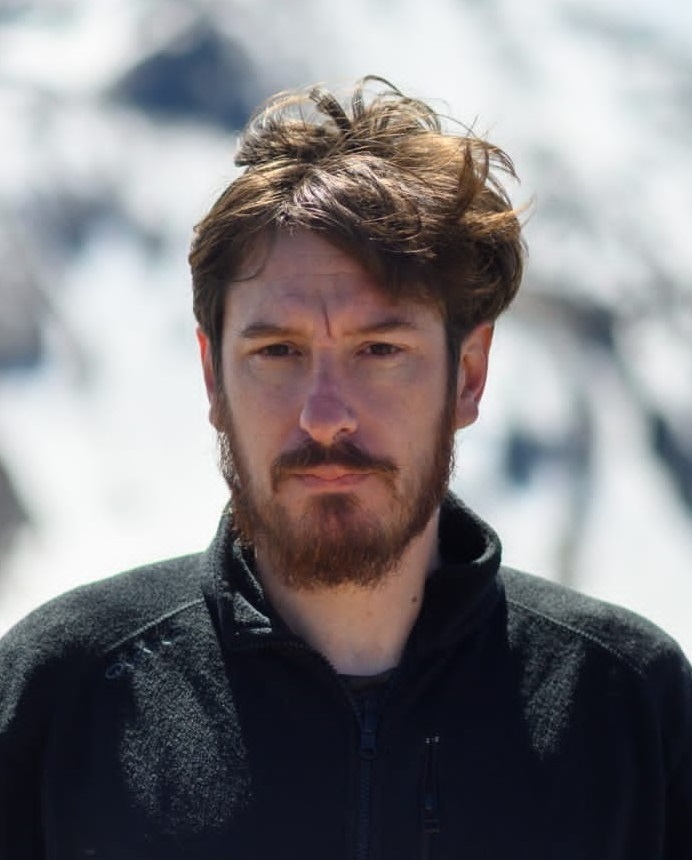}}]{Clement Guilloteau} received his engineering degree (Ms. Eng.) in electronics and signal processing from the National School of Electrotechnics, Electronics, Informatics, Hydraulics and Telecommunication (INP-ENSEEIHT) in Toulouse, France in 2012. He received his Ph.D. degree in atmospheric sciences from the University of Toulouse in 2016. He worked as research assistant for CNRS, France in 2013 and 2016. He has been a postdoctoral scholar in the Department of Civil and Environmental Engineering at the University of California, Irvine from 2017 to 2022 and an associate research specialist since 2023. His research focuses on satellite hydrometeorology.
\end{IEEEbiography}

\begin{IEEEbiography}[{\includegraphics[width=1in,height=1.25in,clip,keepaspectratio]{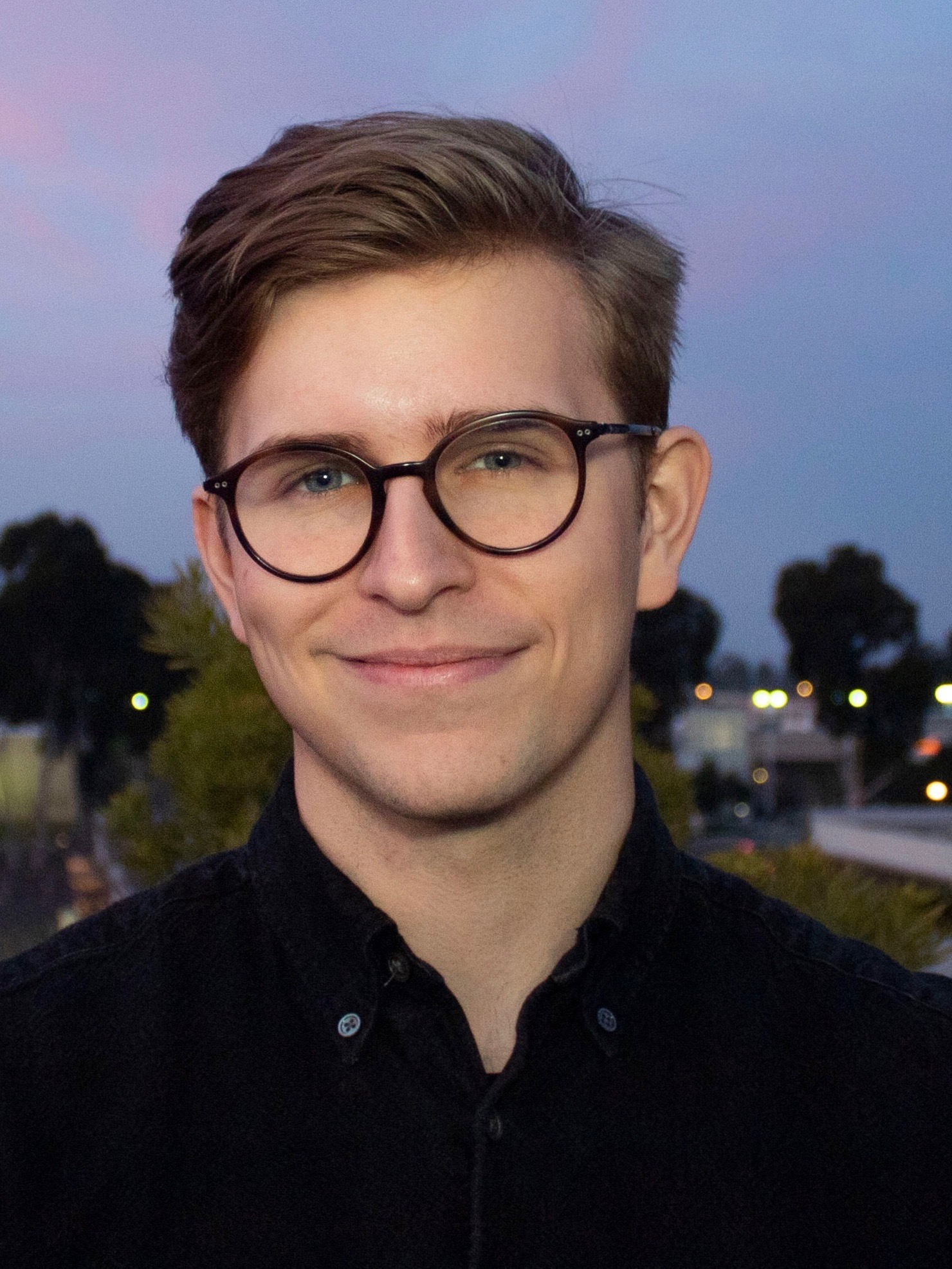}}]{Gavin Kerrigan} received a B.Sc. in mathematics from the Pennsylvania State University in 2019. He is currently a Ph.D. candidate at the University of California, Irvine, where his research focuses on generative models and applications of machine learning in scientific domains.
\end{IEEEbiography}

\begin{IEEEbiography}[{\includegraphics[width=1in,height=1.25in,clip,keepaspectratio]{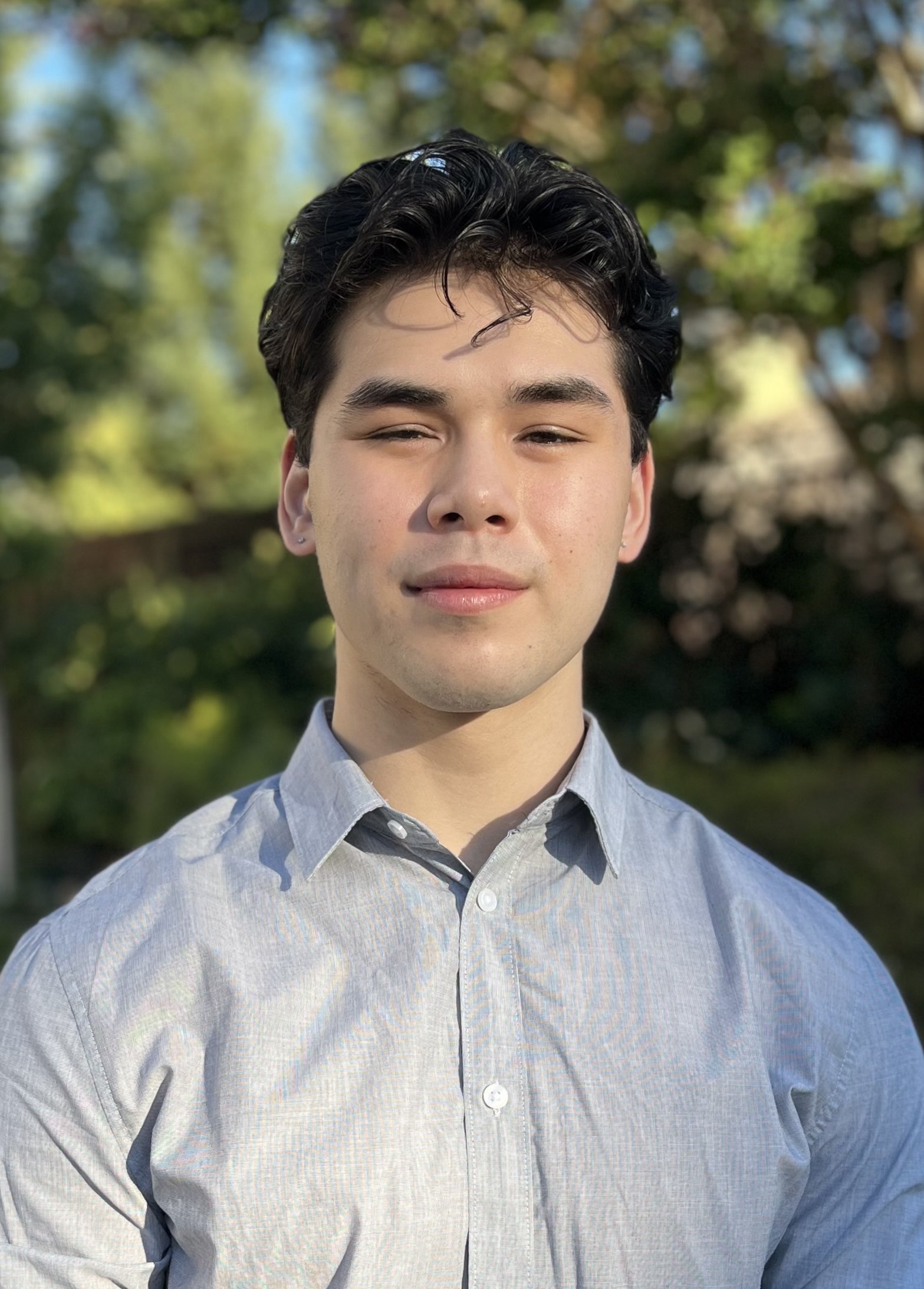}}]{Kai Nelson} is an undergraduate student of Computer Science at the University of California, Irvine. His research interests include deep generative models such as diffusion models, as well as their applications.
\end{IEEEbiography}

\begin{IEEEbiography}
    [{\includegraphics[width=1in,height=1.25in,clip,keepaspectratio]{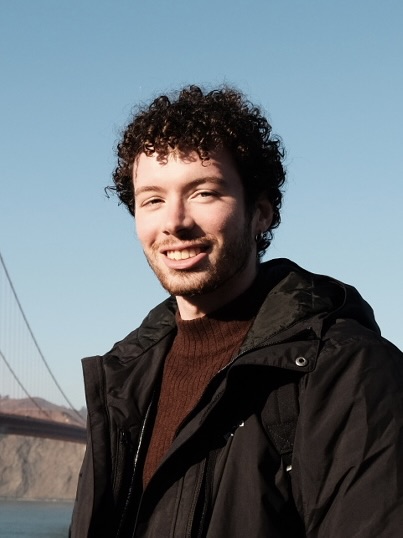}}]{Giosue Migliorini} received his M.Sc. in Data Science from Bocconi University in 2022. Currently, he is a Ph.D. student in the Department of Statistics at the University of California, Irvine. His research interests include Bayesian inference, deep generative models, and applications to spatiotemporal data.
\end{IEEEbiography}

\begin{IEEEbiography}
    [{\includegraphics[width=1in,height=1.25in,clip,keepaspectratio]{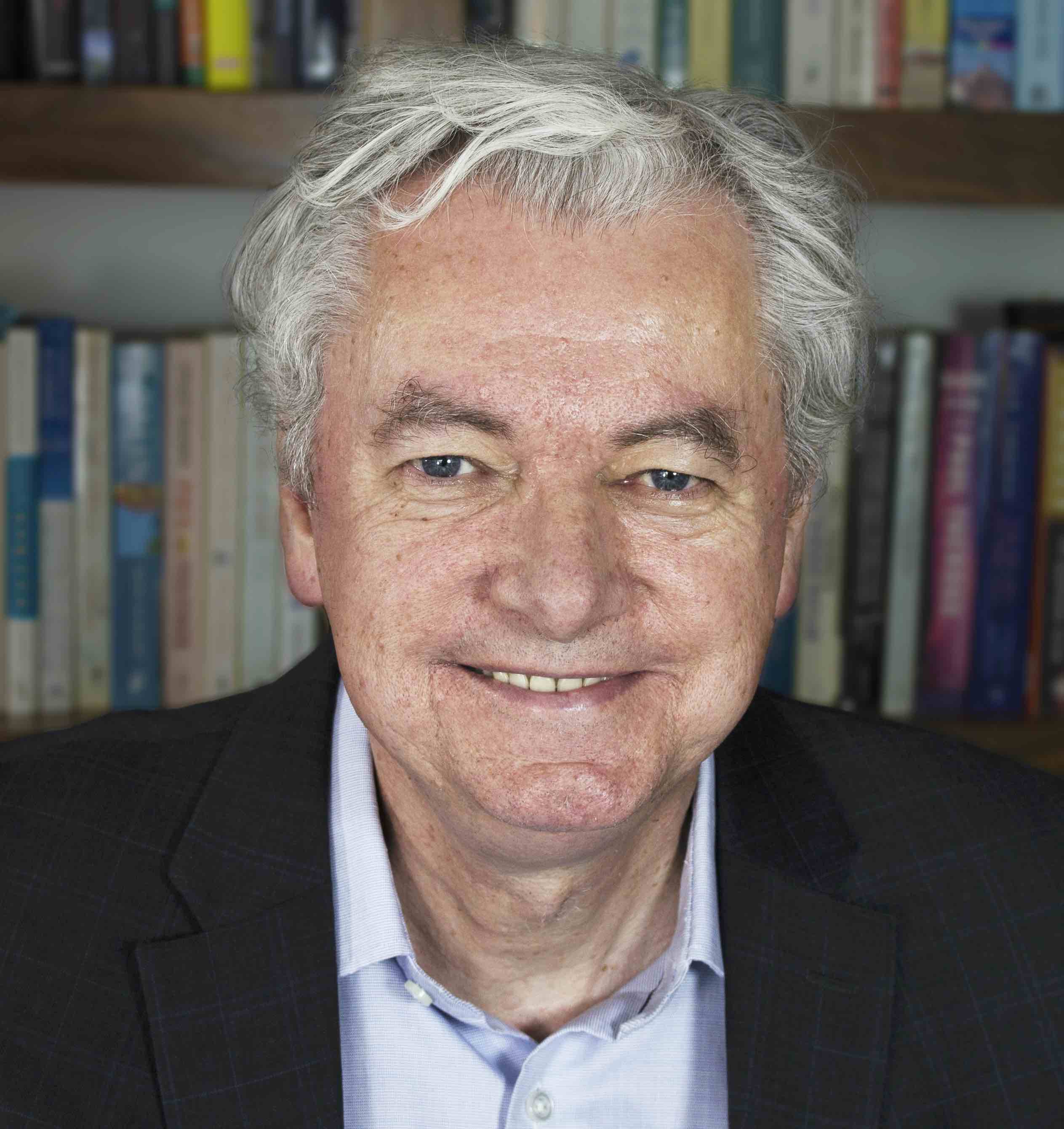}}]{Padhraic Smyth} (Fellow, IEEE) received the B.S. degree in electronic engineering from the National University of Ireland, University College Galway, in 1984 and the M.S. and Ph.D. degrees in electrical engineering from the California Institute of Technology, Pasadena, CA, USA in 1985 and 1988, respectively. From 1988 to 1996 he was a Member of Technical Staff and Technical Group Leader (from 1992) at the  Jet Propulsion Laboratory, California Institute of Technology. Since 1996, he has been on the faculty at the University of California, Irvine, where he is currently a Distinguished Professor    in the Computer Science and Statistics Departments. His research interests include machine learning, pattern recognition, and applied statistics. 
    
Dr. Smyth is a Fellow of the IEEE, of the American Association for the Advancement of Science,  of the Association for Computing Machinery, and of the Association for the Advancement of Artificial Intelligence. He served as program chair of the ACM SIGKDD Conference in 2011 and the Uncertainty in AI Conference in 2013, associate program chair for IJCAI 2022, general chair for AI-Stats 1997, and has served in various senior/area chair positions for conferences such as NeurIPS, ICML, and AAAI. He has also served in editorial and advisory positions for journals such as the Journal of Machine Learning Research, the Journal of the American Statistical Association, and the IEEE Transactions on Knowledge and Data Engineering. He was the founding director of the UCI Center for Machine Learning and Intelligent Systems in 2007, has served as director of the online UCI Machine Learning Repository since 2002, and was academic advisor to Netflix for the Netflix prize competition from 2006 to 2009.
\end{IEEEbiography}

\begin{IEEEbiography}
[{\includegraphics[width=1in,height=1.25in,clip,keepaspectratio]{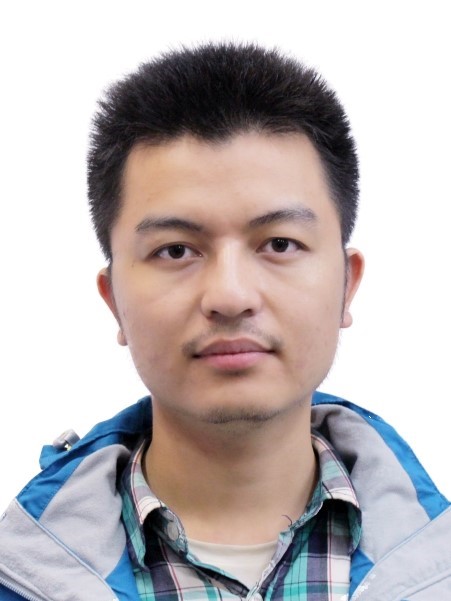}}]
{Runze Li} earned his B.Eng. in Remote Sensing Science and Technology from Wuhan University, China in 2015, and his Ph.D. in Global Environmental Change from Beijing Normal University, China in 2021. Since 2021, he has been a postdoctoral scholar in the Department of Civil and Environmental Engineering at the University of California, Irvine. His research primarily focuses on remote sensing of precipitation.
\end{IEEEbiography}

\begin{IEEEbiography}[{\includegraphics[width=1in,height=1.25in,clip,keepaspectratio]{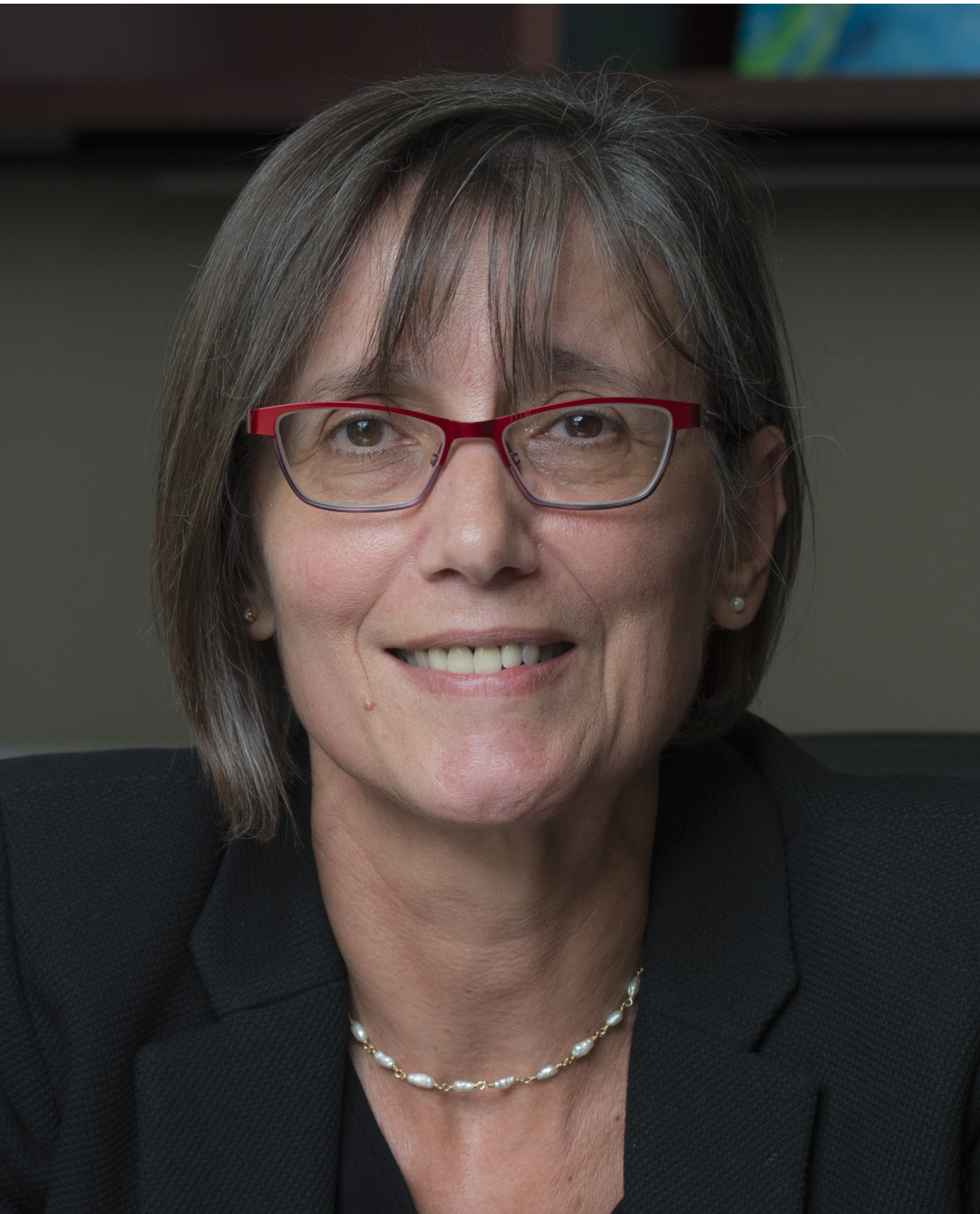}}]{Efi Foufoula-Georgiou} received the Diploma in civil engineering from the National Technical University of Athens, Athens, Greece, in 1979 and the M.Sc. and Ph.D. degrees in environmental engineering from the University of Florida, Gainesville, FL, USA, in 1982 and 1985, respectively. From 1989 to 2016, she was a Distinguished McKnight University Professor in Civil and Environmental Engineering at the University of Minnesota, Minneapolis, MN.  She is currently a Distinguished Professor with the Department of Civil and Environmental Engineering and the Henry Samueli chair at the University of California, Irvine. Her research interests include hydrology, geomorphology and climate dynamics, with emphasis on stochastic modeling and inverse problems. Dr. Foufoula-Georgiou is a Fellow of the American Geophysical Union, American Meteorological Society, and the American Association for the Advancement of Science. She is an elected member of the European Academy of Sciences, the American Academy of Arts and Science and the National Academy of Engineering (NAE). 
\end{IEEEbiography}

\clearpage
\appendix

\subsection{Scoring metrics}
\noindent In the present article we use the Kullback–Leibler divergence (KLD) to assess the similarity between two discretized empirical distributions (histograms). Let us consider a discretized density distribution $D(u)$, where u is the bin index, to be compared to a reference distribution $D_{\textrm{ref}}(u)$. The KLD of $D$ to $D_{\textrm{ref}}$ is defined as
\begin{equation}
    \textrm{KLD}(D,D_{\textrm{ref}})=\sum_{u}D(u)\log\left( \frac{D(u)}{D_{\textrm{ref}}(u)}\right).
\end{equation}

Another metric used in this article is the Brier score. The Brier score is designed to assess the performance of a probabilistic estimation or prediction of a binary variable. Let $y(u) \in \{0,1\}$ be the true state of the predicted variable and $\overline{y}(u) \in [0,1]$ a probabilistic estimate of $y(u)$. The value $\overline{y}(u)$ is the predicted fractional probability of $y(u)$ taking the 1 value, and conversely $1-\overline{y}(u)$ is the predicted fractional probability of $y(u)$ taking the 0 value. The brier score of the probabilistic estimation is computed over $N$ estimations as
\begin{equation}
    \textrm{BS} =\frac{1}{N} \sum_{u=1}^N \left(\overline{y}(u)-y(u)\right)^2.
\end{equation}

\subsection{Data preprocessing}

\noindent Our raw dataset consists of 2473 pairs of precipitation images $\mathbf{x_0}$ and the corresponding MW/IR measurements $\mathbf{z}$ with varying sizes. We preprocess this raw data by first creating a train-test split with 70\% of the pairs used for training and validation, and 30\% of the pairs used for testing. 

After splitting the data, we patch these images, turning these variable-size images into a dataset of images each having a fixed size of $64 \times 64$ pixels. We do this by sliding a $64 \times 64$ window across each image (stacking the $(\mathbf{x_0}, \mathbf{z})$ pairs along the channel dimension) with a stride of $32$ pixels. That is, we begin with the $64 \times 64$ pixel window in the top-left of the image, and iteratively move this window $32$ pixels to the right until any portion of the window would lie outside of the original image. We then move our window back to the left-hand side of the image, but moved down $32$ pixels, followed by iterating the process.

This allows for some overlap between the training images, which can be seen as a form of data augmentation to increase the size of our training dataset. We accept a patch if at least 20\% of the pixels in the patch have a significant amount of rainfall, where a pixel is considered significant if its hourly accumulated rainfall exceeds 0.184 mm, corresponding to the 25th percentile of all pixel values. We perform this preprocessing step as the vast majority of our pixels have zero rainfall, which we found to bias our model toward generating too many zeros in preliminary experiments. However, we note that these requirements for accepting a patch are fairly mild. 

We similarly patch the testing set, but use a stride of 64 to ensure that there is no overlap in the test-set images. This is done to ensure that our evaluation is not overly optimistic and that we are not evaluating on the same images more than once. This patching procedure resulted in a total of 3617 $64 \times 64$ images for training and 486 $64 \times 64$ images for testing.

After patching the data, we normalize the training data by transforming each pixel into the range $[-1, 1]$. We do this by computing the minimum and maximum value of each channel across the training set (i.e. independently for the accumulated rainfall, and independently for each of the IR/MW channels), followed by channel-wise centering and re-scaling by these values. At testing time, the images from the model are generated on this scale, followed by un-normalizing the generated images back to their natural scale.

\subsection{Training details}

\noindent As our dataset is relatively small, we perform additional data augmentation at training time. When we sample a datapoint $(\mathbf{x_0}, \mathbf{z})$ from the training set, we potentially flip this image around its vertical and/or horizontal axes. These flips both occur with an independent probability of $0.5$. We note that data augmentation is a standard technique in computer vision \cite{yang2022image} which is widely applied, especially in settings with limited amounts of training data. 

All models were trained using Adam \cite{kingma2014adam} for a maximum of $750$ epochs with early stopping, using the validation loss as a stopping criterion. We train with a batch size of 64 and perform a grid-search to determine the learning rate and weight decay, leaving all other optimizer values as their defaults suggested by the original paper \cite{kingma2014adam}. Our final model was trained with a learning rate of $10^{-4}$ and weight decay of $0$. All of our models are trained on a single NVIDIA A5000 GPU with 24GB VRAM. On our hardware, each complete training run requires roughly $6$ hours of wall-clock time. At testing time, sampling an ensemble of $128$ images for a given MW/IR image $\mathbf{z}$ required roughly $2$ minutes of wall-clock time. 

\subsection{Diffusion details}

\noindent We found in preliminary experiments that $K=1000$ diffusion steps provided a balance between sampling efficiency and quality, and left this value fixed throughout our model selection process. 

We performed a grid search over noise schedules, and our final model uses the sigmoid noise scheduler proposed in \cite{jabri2023scalable}. Intuitively, this noise schedule has the appeal of slowing down the diffusion process near the initial iterations, which are the most crucial steps in the generative process. We refer to Algorithm 4 of \cite{jabri2023scalable} for a precise description.

In addition, our final model uses the $v$-parametrization proposed by Salimans et al. \cite{salimans2022progressive}. This is a model parametrization which is an alternative to the $\epsilon$-parametrization in Equation \eqref{eqn:eps_to_mu} that we found to give better results in preliminary experiments. This parametrization is motivated by the desire to stabilize the model predictions as the signal-to-noise-ratio varies throughout the diffusion process.

In some more detail, given noise $\boldsymbol{\epsilon}$ and a clean image $\mathbf{x_0}$ , we may define $\mathbf{v_k}  = \sqrt{\overline{\alpha}_k} \boldsymbol{\epsilon} - \sqrt{1 - \overline{\alpha_k}}\mathbf{x_0}$ as a linear combination of the noise and clean image. Training a model $v_{\boldsymbol{\theta}}(k, \mathbf{x_k}, \mathbf{z})$ on this objective amounts to minimizing the expected MSE between the model’s output (given the noisy $\mathbf{x_k}$ and $\mathbf{z}$) and the objective $\mathbf{v_k}$. That is, our loss function is simply
\begin{equation}
    \E_{\mathbf{x_0}, \mathbf{z}, \boldsymbol{\epsilon}, k}\left[ \norm{\mathbf{v_k} - v_{\boldsymbol{\theta}}(k, \mathbf{x_k}, \mathbf{z})}^2 \right]
\end{equation}

which is analogous to the loss show in Equation \eqref{eqn:eps_loss}.

We emphasize that this is equivalent to the $\boldsymbol{\epsilon}$-prediction setup, as one may recover $\epsilon_{\boldsymbol{\theta}}(k, \mathbf{x_k}, \mathbf{z})$ from $v_{\boldsymbol{\theta}}(k, \mathbf{x_k}, \mathbf{z})$ via
\begin{multline*}
    \epsilon_{\boldsymbol{\theta}}(k, \mathbf{x_k}, \mathbf{z}) = \\
    \frac{1}{\sqrt{\overline{\alpha}_k^{-1} - 1}} \left( \left( \overline{\alpha}_k^{-1} - \sqrt{\alpha_k} \right) \mathbf{x_k} - \sqrt{1 - \overline{\alpha}_k} v_{\boldsymbol{\theta}}(k, \mathbf{x_k}, \mathbf{z}) \right).
\end{multline*}

We refer to \cite{salimans2022progressive} for additional details.

\subsection{Architecture details}

Our architecture consists of two UNets \cite{ronneberger2015u} stacked in sequence. We call the first UNet the predictive network, which can intuitively be thought of as predicting a coarse estimate of the conditional mean. The second UNet, which we call the generative network, can be viewed as adding additional fine-grained details to this prediction.

The full model takes in a noisy version of the rainfall image $\mathbf{x_k}$ along with the conditioning images $\mathbf{z}$ and timestep $k$. We apply standard sinusoidal positional embeddings to encode the diffusion iteration $k$ as a $16$-dimensional vector \cite{vaswani2017attention}, followed by a small feed-forward network consisting of a linear transformation to $64$ dimensions, a GELU activation \cite{hendrycks2016gaussian}, and a final linear transformation to 64 dimensions. This small network is not shared across the predictive and generative UNets.

The UNet architecture for the predictive network is standard and adapted from \cite{wang2024ddpm}. This architecture consists of several residual blocks, followed by downsampling in the first half of the network and upsampling in the second half of the network. The time embedding is passed to each block of the network, and is used to produce a shift and scale factor for the output feature map of each block via a SiLU activation and a linear layer. The generative network generally follows the same architecture, with the key differences described below. No parameters are shared across these two UNets.

To perform a forward pass with our model, we first take only the conditioning information $\mathbf{z}$ and diffusion step $k$ and pass these through the predictive UNet. Since this predictive UNet does not take in the noisy version of the image $\mathbf{x_k}$, we may see the output of this predictive UNet as a prediction of the conditional mean. The output of the predictive UNet is a $64 \times 64$ single channel image. We also return the intermediate feature maps from the predictive UNet which are later used as inputs to the generative UNet.

Then, we compute a residual, which is the pixelwise difference between the noisy image $\mathbf{x_k}$ and the output of the predictive UNet. This residual, along with the output of the predictive UNet and the conditioning images, are concatenated channelwise before being passed into the generative UNet along with the diffusion step $k$ and intermediate feature maps from the predictive UNet. In the forward pass of the generative UNet, the corresponding intermediate feature maps from the predictive UNet are channelwise concatenated with the feature maps of the previous layer of the generative UNet. Lastly, the output of the generative UNet is combined additively with the output of the predictive UNet to produce the final output. 

\end{document}